\documentclass[12pt]{article}
\usepackage{geometry}
\usepackage{amssymb}
\usepackage{amsmath}
\usepackage{amsthm}
\usepackage{mathtools}
  
\usepackage{amsfonts}
\usepackage{graphicx}
\usepackage{enumerate}
\usepackage{color}
\usepackage{hyperref}
\usepackage[onehalfspacing]{setspace}
\usepackage[longnamesfirst]{natbib}
\usepackage{subfigure}
\usepackage{fullpage}

\definecolor{MyDarkBlue}{rgb}{0,0.08,0.45}
\hypersetup{linkcolor=MyDarkBlue, citecolor=MyDarkBlue, colorlinks=true}

\newtheorem{theorem}{Theorem}

\newtheorem{lemma}{Lemma}

\newtheorem{fact}{Fact}

\theoremstyle{definition}
\newtheorem{definition}{Definition}

\newcommand{\finexhere}{\begingroup \let\mathqed\math@qedhere
    \let\qed@elt\setQED@elt \blacktriangle@stack\relax\relax \endgroup}
\DeclareMathOperator*{\argmax}{arg\,max}

\begin{document}

\title{Stochastic Choice and Optimal Sequential Sampling\thanks{%
We thank Stefano DellaVigna and Antonio Rangel for very stimulating conversations, 
Ian Krajbich, Carriel Armel and Antonio Rangel for sharing their experimental data with us,
In Young Cho and
Jonathan Libgober for expert research assistance, the Sloan Foundation and
NSF grants SES-0954162, SES-1123729, and CAREER grant SES-1255062 for
support, and seminar audiences at the ASSA annual meetings, Behavioral Game
Theory conference at UCL; economics seminars at Chicago, Harvard/MIT, Northwestern, Queen's, SITE, Stanford GSB, and Toronto; and a seminar at the  Computational Cognitive Science Group at MIT for useful comments.}}
\author{Drew Fudenberg\thanks{%
Department of Economics, Harvard University} \and Philipp Strack\thanks{%
Department of Economics, University of California, Berkeley} \and Tomasz
Strzalecki\thanks{%
Department of Economics, Harvard University}}
\maketitle

\begin{abstract}
We model the joint distribution of choice probabilities and decision times
in binary choice tasks as the solution to a problem of optimal sequential
sampling, where the agent is uncertain of the utility of each action and
pays a constant cost per unit time for gathering information. In the
resulting optimal policy, the agent's choices are more likely to be correct
when the agent chooses to decide quickly, provided that the agent's prior
beliefs are correct. For this reason it better matches the observed
correlation between decision time and choice probability than does the
classical drift-diffusion model, where the agent is uncertain which of two
actions is best but knows the utility difference between them.
\end{abstract}

\newpage

\section{Introduction}

In laboratory experiments where individuals are repeatedly faced with the
same choice set, the observed choices are stochastic---individuals don't
always choose the same item from a given choice set, even when the choices
are made only a few minutes apart.\footnote{%
See \cite{Hey95,Hey01}, \cite{Ballingeretal}, \cite{Cheremukhinetal}.} In
addition, individuals don't always take the same amount of time to make a
given decision---response times are stochastic as well. Our goal here is to
model the joint distribution of choice probabilities and decision times in
choice tasks, which we call a \emph{choice process}.

We restrict attention to the binary choice tasks that have been used in most
neuroscience choice experiments, and suppose that the agent is choosing
between two items that we call left $(l)$ and right $(r).$ In this setting
we can ask how the probability of the more frequent (i.e., modal) choice
varies with the time taken to make the decision. If the agent is learning
during the decision process, and is stopped by the experimenter at an
exogenous time, we would expect the data to display a \emph{speed-accuracy
tradeoff, }in the sense that the agent makes more accurate decisions
(chooses the modal object more often) when given more time to decide.
However, in many choice experiments there is instead the opposite
correlation: slower decisions are less likely to be modal %
\citep{Swensson72,Luce86,Ratcliff-review}.

To explain this, we develop a new variant of the drift diffusion model
(DDM); other versions of the DDM have been extensively applied to choice
processes in the neuroscience and psychology literatures.\footnote{%
The DDM was first proposed as a model of choice processes in perception
tasks, where the subjects are asked to correctly identify visual or auditory
stimuli. (For recent reviews see \cite{Ratcliff-review} and \cite%
{Shadlen-review}.) More recently, DDM-style models have recently been
applied to choice tasks, where subjects are choosing from a set of
consumption goods presented to them. \cite{ClitheroRangel13, Krajbichetal10,
Krajbichetal11, Krajbichetal12, Milosavljevicetal10, Reutskajaetal11}} The
specification of a DDM begins with a diffusion process $Z_{t}$ that
represents information the subjects is receiving over time, and two disjoint
stopping regions $S_{t}^{l}$ and $S_{t}^{r}$. The agent stops if at some
point in time $t$ it happens that $Z_{t}\in S_{t}^{l}$ (in which case he
chooses $l$) or $Z_{t}\in S_{t}^{r}$ (in which case he chooses $r$);
otherwise the agent continues. Because the evolution of the diffusion
depends on which choice is better, the model predicts a joint probability
distribution on choices and response times conditional on the true state of
the world, which is unknown to the agent.

The oldest and most commonly used version of the DDM (which we will refer to
as \emph{simple} DDM) specifies that the stopping regions are constant in
time, i.e., $S_{t}^{l}=S^{l}$ and $S_{t}^{r}=S^{r}$, and that $Z_{t}$ is a
Brownian motion with drift equal to the difference in utilities of the
items. This specification corresponds to the optimal decision rule for a
Bayesian agent who believes that there are only two states of the world
corresponding to whether action $l$ or action $r$ is optimal, pays a
constant flow cost per unit of time, and at each point in time decides
whether to continue gathering the information or to stop and take an action.%
\footnote{\cite{Wald47} stated and solved this as a hypothesis testing
problem; \cite{ABG} solved the corresponding Bayesian version. These models
were brought to the psychology literature by \cite{Stone60} and \cite%
{Edwards65}.} The constant stopping regions of the simple DDM\ imply that
the expected amount of time that an agent will gather information depends
only on the current value of $Z_{t},$ and not on how much time the agent has
already spent observing the signal process, and that the probability of the
modal choice is independent of the distribution of stopping times.\footnote{%
\cite{Stone60} proved this independence directly for the simple DDM in
discrete time. Our Theorem \ref{thm:monot} shows that the independence is a
consequence of the stopping boundaries being constant.} In contrast, in many
psychological tasks \citep{Churchlandetal, Ditterich06} reaction times tend
to be higher in incorrect than correct trials. For this reason, when the
simple DDM is applied to choice data, it predicts response times that are
too long for choices in which the stimulus is weak, or the utility
difference between them is small. Ad-hoc extensions of DDM have been
developed to better match the data, by allowing more general processes $%
Z_{t} $ or stopping regions $S_{t}^{i}$, see e.g., \cite{Laming68,
LinkHeath75, Ratcliff78}. However, past work has left open the question of
whether these generalizations correspond to any particular learning problem,
and if so, what form those problems take.

Our main focus in this paper is to provide learning-theoretic foundations
for an alternative form of DDM, where the agent's behavior is the solution
to a sequential sampling problem with a constant cost per unit time as in
the simple DDM but with a prior that allows her to learn not only which
alternative is better, but also by how much. In this \emph{%
uncertain-difference DDM}, the state of the world is the vector $\theta
=(\theta ^{l},\theta ^{r})$ of the utilities of the two choice. In this
model an agent with a large sample and $Z_{t}$ close to zero will decide the
utility difference is small, and so be more eager to stop than an agent with
the same $Z_{t}$ but a small sample.

Our main insight is that the nature of the learning problem matters for the
optimal stopping strategy and thus for the distribution of choices and
response times. In particular, we show that in the uncertain-difference DDM
it is optimal to have the range of $Z_{t}$ for which the agent continues to
sample collapse to $0$ as time goes to infinity, and moreover that is does
so asymptotically at rate $1/t$. The intuition for the fact that the
boundary should converge to 0 is not itself new, and has been put forward
both as a heuristic in various related models and as a way to better fit the
data \citep[see, e.g.,][]{ShadlenKiani13} but we provide the first precise
statement and solution of an optimization problem that generates decreasing
boundaries, thus providing a foundation for their use in empirical work,
such as the exogenous exponentially-declining boundaries in \cite%
{Milosavljevic10}.\footnote{\cite{Drugowitsch12} state a decision problem
with two states (e.g. known payoff difference), a fixed terminal time, and
time-dependent cost functions, and discuss how to use dynamic programming to
numerically compute the solution. Note that even with constant costs the
boundary is decreasing when the horizon is finite.} We then use
approximation results and numerical methods to determine the functional form
of the boundary, thus providing guidance about what the rate of collapse
might be expected to be.

Finally, we investigate the consequences of allowing the flow cost to
vary arbitrarily with time. Intuitively, if the cost decreases quickly enough,
this might outweigh the diminishing effectiveness of learning and lead to an
increasing boundary. We show that this intuition is correct, and more
strongly that any stopping region at all can be rationalized by a suitable
choice of a cost function. Thus optimal stopping on its own imposes
essentially no restrictions on the observed choice process, and so it is
compatible with the boundaries assumed in \cite{Ratcliff-review} and \cite%
{ShadlenKiani13}. However, the force of the model derives from its joint
assumptions about the evolution of beliefs and the cost function, and the
cost functions needed to rationalize various forms of boundary may or may
not seem plausible in the relevant applications.

One motivation for modeling the joint distribution of decision times and
choices is that the additional information provided by decision times can
lead to models that are closer to the underlying neural mechanisms and may
therefore be more robust.\footnote{%
See \cite{ShadlenKiani13} and \cite{Bogaczetal06} for discussions of how
DDM-type models help explain the correlation between decision times and
neurophysiological data such as neuronal firing rates.} In addition, as
shown by \cite{ClitheroRangel13}, including decision times leads to better
out-of-sample predictions of choice probabilities. In other settings than
the simple choice tasks we study here, decision times can been used to
classify decisions as \textquotedblleft
instinctive/heuristic\textquotedblright\ or \textquotedblleft
cognitive,\textquotedblright\ as in \cite{Rubinstein07} and \cite{Randetal12}%
.

In addition to the papers cited above, our theoretical approach is closely
related to the recent work of \cite{Woodford}. In his model the agent can
optimize the dependence of the process $Z_{t}$ on $\theta $ subject to a
Shannon capacity constraint, but the stopping rule is constrained to have
time-invariant boundaries. In our model the process $Z_{t}$ is exogenous but
the stopping rule is optimal subject to a cost, so the two approaches are
complementary: both models feature an optimization problem, but over
different spaces.

\cite{GL} and \cite{KSVB} look at decisions derived from optimal stopping
rules where the gains from sampling are exogenously specified as opposed to
being derived from Bayesian updating, as they are here; neither paper
examines the correlation between decision time and accuracy. \cite%
{Oneanddone} studies the optimal predetermined sample size for an agent
whose cost of time arises from the opportunity to make future decisions;
they find that for a range of parameters the optimal sample size is one.

\cite{Natenzon} and \cite{Lu} study models with an exogenous stopping rule.
They treat time as a parameter of the choice function, and not as an
observable in its own right. Similarly, decision field theory \citep{DFT}
specifies an exogenous stopping rule; while this literature does discuss the
effect of time pressure it does not treat the observed choice times as data.
Our model makes joint predictions about decisions and response times because
the stopping time is chosen optimally. These additional predictions provide
more structure on stochastic choice and can help us develop more accurate
models.

\section{Choice Processes and DDMs}

\label{sec:DDM}

\subsection{Observables}

Let $A=\{l,r\}$ be the set of alternatives, which we will call left $(l)$
and right $(r).$ Let $T=[0,+\infty )$ be the set of decision times---the
times at which the agent is observed to state a choice. The analyst observes
a joint probability distribution $P\in \Delta (A\times T)$; we call this a 
\emph{choice process}. We will decompose $P$ as 
\begin{equation*}
p^{i}(t)\text{ and }F(t)
\end{equation*}%
where $p^{i}(t)$ is probability of choosing $i\in A$ conditional on stopping
at time $t$ and $F(t)=P(A\times \lbrack 0,t])$ is the cdf of the marginal
distribution of decision times.\footnote{%
Formally, we assume that $P$ is a Borel probability measure on $A\times T$.
The conditional probabilities $p^{i}(t)$ exist by Theorems 10.2.1 and 10.2.2
of \cite{Dudley}.}

It will also be useful to decompose $P$ the other way 
\begin{equation*}
P^{i}\text{ and }F^{i}(t)
\end{equation*}%
where $P^{i}=P(\{i\}\times T)$ is the overall probability of choosing $i\in
A $ at any time, and $F^{i}(t)=\frac{P(\{i\}\times [0, t])}{P^i}$ is the cdf
of time conditional on choosing $i\in A$.

\subsection{Speed and Accuracy}

It is easy to define accuracy in perceptual decision tasks, since in such
settings the analyst knows which option is `correct.' However, in choice
tasks the agents' preferences are subjective and may be unknown to the
researcher.\footnote{%
In some cases the analyst has a proxy of the preference in form of a
separately elicited ranking, see, e.g., \cite{Krajbichetal10}, \cite%
{Milosavljevic10}, \cite{Krajbichetal12}.} One way of defining accuracy is
with respect to the modal choice, as we expect that the objects the agent
chooses more often are in some sense ``better;'' we denote the modal choice
by $m$, the other one by $o$.

The simplest possible relationship between choices and times is no
relationship at all, that is when the distribution of stopping times is
independent of the distribution of choices. We will equivalently define this
property as follows.

\begin{definition}
$P$ displays a \emph{speed-accuracy independence} iff $p^{m}(t)$ is a
constant function of $t$
\end{definition}

Speed-accuracy independence is a necessary implication of the simple DDM,
which we introduce formally in the next section. The case of independence is
by nature very knife-edge, and is natural to relax it. In this paper, we
focus on qualitatively capturing a positive or a negative
correlation between choices and time. To do this, we introduce the following
definition.

\begin{definition}
$P$ displays a \emph{speed-accuracy tradeoff} iff $p^{m}(t)$ is an
increasing function of $t$.
\end{definition}

Note that this definition requires that the tradeoff holds for all times $t$%
. We expect there to be a speed-accuracy tradeoff if the agent is learning
about the options before her and is stopped by the experimenter at an
exogenous stochastic time, as by waiting longer he obtains more information
and can make more informed choices. But even if the agent is learning, the
observed choice process $P$ need not display a choice-accuracy tradeoff if
the stopping time is not exogenous but chosen by the agent as a function of
what has been learned so far. In this case, the agent might stop sooner when 
she thinks his choice is likely to be accurate, so the choice of a
stopping time may push towards the opposite side of the speed-accuracy
tradeoff.



\begin{definition}
\label{def:sat} $P$ displays a \emph{speed-accuracy complementarity} iff $p^{m}(t)$
is a decreasing function of $t$.
\end{definition}

A priori we could observe both kinds of $P$, perhaps depending on the
circumstances. This is indeed the case; for example, \cite{Swensson72} and 
\cite{Luce86} report that speed-accuracy complementarity is observed under
normal conditions, but speed-accuracy tradeoff is observed when subjects are
incentivized on speed; see also, \cite{ShadlenKiani13}.\footnote{%
This could be explained if agents are stopping at random under time
pressure, but are using some other rule under normal circumstances; for
example they are following the uncertain-difference DDM, to be described in
Section \ref{sec:ud}.}

The speed-accuracy tradeoff can be equivalently expressed in terms of the
monotone likelihood ratio property. Let $P$ be a choice process and let $%
f^{i}$ be the density of $F^{i}$ with respect to $F$. Suppose that $F^{m}$
is absolutely continuous w.r.t. $F^{o}$; we say that $F^{m}$ and $F^{o}$
have the \emph{monotone likelihood ratio property}, denoted $F^{m}\succsim _{%
\text{MLRP}}F^{o}$, if the likelihood $f^{m}(t)/f^{o}(t)$ is an increasing
function.

Though the above concepts will be useful for theoretical analysis, in
empirical work time periods will need to be binned to get useful test
statistics. For this reason we introduce two weaker concepts that are less
sensitive to finite samples, as their oscillation is mitigated by
conditioning on larger sets of the form $[0,t]$. First, let $Q^{i}(t):=\frac{%
P(\{i\}\times \lbrack 0,t])}{F(t)}$ be the probability of choosing $i$
conditional on stopping in the interval $[0,t]$. Second, we say that $F^{m}$
first order stochastically dominates $F^{o}$, denoted $F^{m}\succsim _{\text{%
FOSD}}F^{o}$ if $F^{m}(t)\leq F^{o}(t)$ for all $t\in T$. Below, we
summarize the relationships between these concepts.

\begin{fact}
\label{fact:FOSD} \ 

\begin{enumerate}
\item Let $P$ be a choice process and suppose that $F^m$ is absolutely
continuous w.r.t. $F^o$. Then $P$ displays the speed-accuracy tradeoff
(complementarity/independence) if and only if $F^m\succsim_{\text{MLRP}}F^o$
($F^m\precsim_{\text{MLRP}}F^o$ / $F^m=F^o$).

\item If $P$ displays a speed-accuracy tradeoff (complementarity,
independence), then $Q^{m}(t)$ is an increasing (decreasing, constant)
function

\item If $Q^{m}(t)$ is an increasing (decreasing, constant) function, then $%
F^{m}\succsim _{\text{FOSD}}F^{o}$ ($F^{m}\precsim _{\text{FOSD}}F^{o}$, $%
F^{m}=F^{o}$).
\end{enumerate}
\end{fact}

\begin{figure}[p!]
\begin{center}
\includegraphics[width=\textwidth]{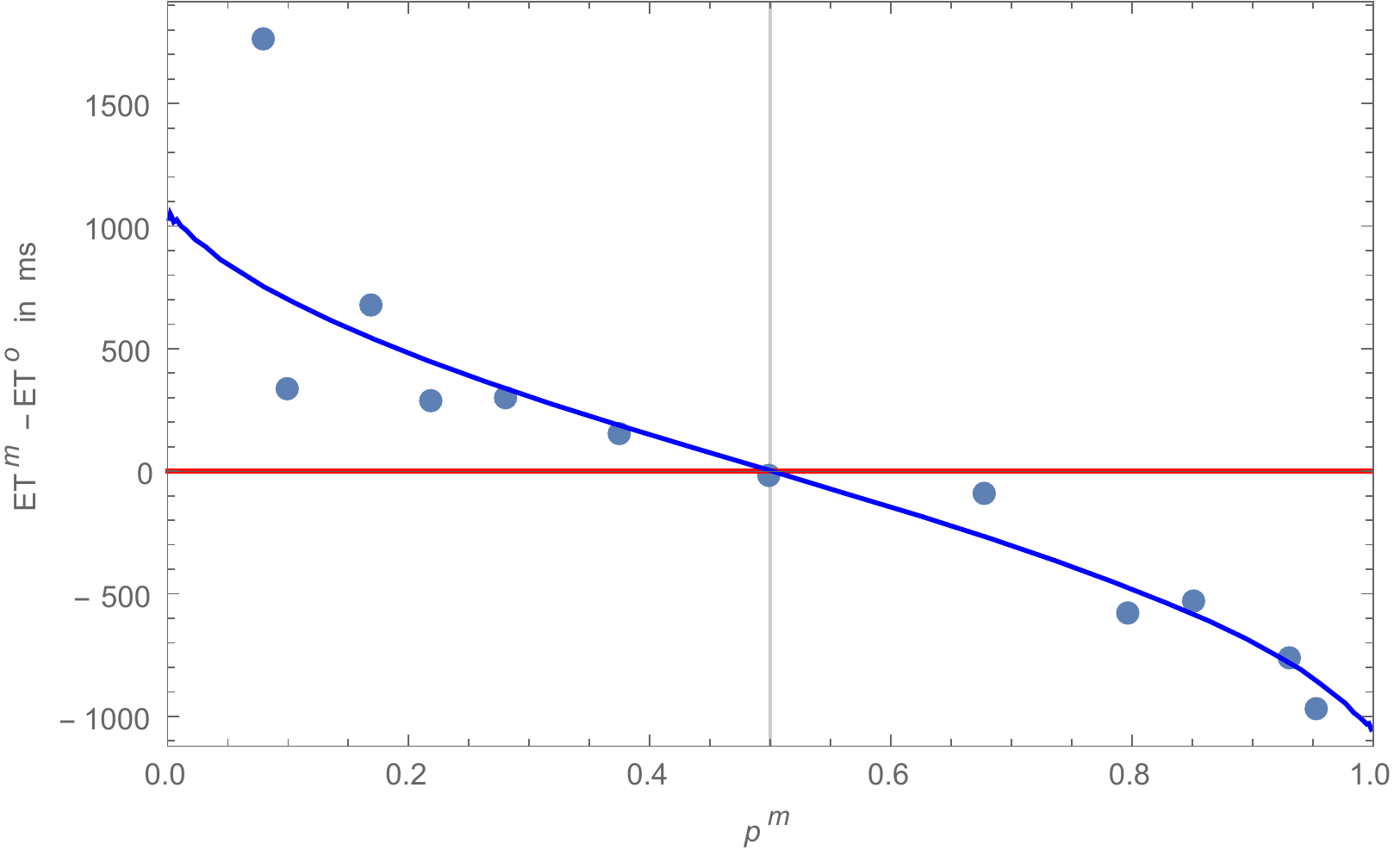}
\end{center}
\caption{The dots display the data from an experiment by \protect\cite%
{Krajbichetal10}. (Since each alternative pair is sampled only once, we used
self-reported rankings collected by \protect\cite{Krajbichetal10} to cluster
the budget sets and compute choice probabilities.) The blue line displays
simulations of the approximately optimal barrier $\bar b$ for the
uncertain-difference DDM (derived in Section \protect\ref{sec:ud}) with
parameters $c=0.05,\protect\sigma _{0}=\protect\alpha =1$; each simulation
uses $5\cdot 10^{6}$ draws and time is discretized with $dt=0.01$. The
expected time to choose $l$ is lower than the expected time to choose $r$
for pairs of alternatives where $l $ is the modal choice; it is higher for
pairs where $r$ is modal. The red line displays predictions of the simple
DDM for any combination of parameter values.}
\label{fig:1}
\end{figure}

In particular, under speed-accuracy complementarity, the expected time to
choose the modal alternative is smaller than the expected time to choose the
other alternative, i.e., $Et^{m}:=\int_{0}^{\infty }tdF^{m}(t)\leq
\int_{0}^{\infty }tdF^{o}(t)=:Et^{o}$. Figure \ref{fig:1} shows that this
fits the experimental data of \cite{Krajbichetal10}.

The proof of Fact 1 and all other results is in the Appendix. In the next
section of the paper we study a particular class of choice processes $P$,
called \emph{drift-diffusion models} (DDM). Such models are defined by a
signal process $Z_{t}$ and a stopping boundary $b_{t} $. We characterize the
three subclasses of DDM: speed-accuracy tradeoff, complementarity, and
independence in terms the function $b_{t}$.

\subsection{DDM representations}

DDM representations have been widely used in the psychology and neuroscience
literatures \citep{Ratcliff-review,Shadlen-review,FehrRangel}. The two main
ingredients of a DDM are the stimulus process $Z_{t}$ and a boundary $b_{t}$.

The stimulus process $Z_{t}$ is assumed to be a Brownian motion with drift $%
\delta $ and variance~$\alpha ^{2}$: 
\begin{equation}
Z_{t}=\delta t+\alpha B_{t},  \label{eq:Z}
\end{equation}%
where $B_{t}$ is a standard Brownian motion. In early applications of DDM
such as \cite{Ratcliff78} $Z_{t}$ was is not observed by the experimenter.
In some recent applications of DDM to neuroscience, the analyst may observe
signals that are correlated with~$Z_{t}$; for example the neural
firing rates of both single neurons \citep{HanesSchall96} and populations of
them \citep[e.g.,][]{Ratcliffetal03}. In the later sections we interpret the
process $Z_{t}$ as a signal about the utility difference between the two
alternatives.

Suppose that the agent stops at a fixed time $s$ and choose $l$ if $Z_{s}>0$
and $r$ if $Z_{s}<0$ (and flip a coin if there is a tie). Let $p^{m}(s)$ be
the frequency of the modal choice. It is easy to see that if $\delta >0$
then the modal choice is $l$, and its probability $p^{m}(s)$ is an
increasing function of the exogenous stopping time $s$: the process starts
at $Z_{0}=0$, so if $s=0$ each choice is equally likely. At each subsequent
point in time $s>0$ the distribution of $Z_{s}$ is $\mathcal{N}(\delta
s,\alpha ^{2}s)$; thus, the probability of $Z_{s}>0$ increases. The same
happens if the agent stops at a stochastic time $\tau $ that is independent
of the process $Z_{t}$. Thus, an \emph{exogenously} stopped process $Z_{t}$
leads to a speed-accuracy tradeoff. We will now see that if the process $%
Z_{t}$ is stopped \emph{endogenously}, i.e., depending on its value, then
this effect can be reversed.

The canonical example of a stopping time that depends on $Z_{t}$ is the
hitting time of a boundary. A \emph{boundary} is a function $b:\mathbb{R}%
_{+}\rightarrow \mathbb{R}$. Define the hitting time $\tau $ 
\begin{equation}
\tau =\inf \{t\geq 0:|Z_{t}|\geq b_{t}\},  \label{eq:tau}
\end{equation}%
i.e., the first time the absolute value of the process $Z_{t}$ hits the
boundary. Let $P(\delta ,\alpha ,b)\in \Delta (A\times T)$ be the choice
process induced by $\tau $ and a decision rule that chooses $l$ if $%
Z_{t}=b_{t}$ and $r$ if $Z_{t}=-b_{t}$.\footnote{%
There are boundaries for which there is a positive probability that $\tau
=\infty $. This cannot happen for the primitive objects that we consider
here, which are choice processes. Thus, we only focus on those boundaries
that lead the agent to stop in finite tine with probability 1. This property
will be satisfied in any model where the stopping time comes from a
statistical decision problem in which never stopping incurs an infinite cost
and the value of full information is finite.}

\begin{definition}
A choice process $P$ has a \emph{DDM representation} ($\delta ,\alpha ,b$)
if $P=P(\delta ,\alpha ,b)$. \footnote{%
Note that the parameter $\alpha $ can be removed here by setting $%
\alpha^{\prime }=1,\delta ^{\prime }=\delta /\alpha ,$ and $b^{\prime
}=b/\alpha$. By a similar argument, $\delta$ can be assumed to be $-1, 0,$
or $1$. We nonetheless retain $\alpha$ and $\delta$ here as we will use them
in the next section to distinguish between utility and signal strength.}
\end{definition}

We note that the assumption that the process $Z_{t}$ is Brownian is an
important one, as without it the model is vacuous.

\begin{fact}
\label{fact:DDM} Any choice process $P$ has a \textquotedblleft
DDM-like\textquotedblright\ representation where the stochastic process $%
Z_{t}$ is arbitrary.
\end{fact}

In particular an arbitrary $P$ may correspond to a process $Z_{t}$ with
jumps (or a filtration that is not right-continuous). However, the standard
assumption in the literature is that $Z_{t}$ is Brownian.\footnote{%
For the Brownian motion \cite{Smith00} and \cite{Peskir02} show that the
distribution of hitting times satisfies a system of integral equations
depending on the boundary. The inverse problem of finding a boundary such
that the first hitting time has a given distribution was studied in \cite%
{iscoe1999integrated} in the case Brownian Random walks and they propose a
Monte Carlo algorithm for solving it numerically.} The next result
characterizes the relationship between speed and accuracy in the class of
choice processes that admit a DDM representation.

\begin{theorem}
\label{thm:monot} Suppose that $P$ has a DDM representation $(\alpha,
\delta, b)$. Then $P$ displays a speed-accuracy tradeoff (complementarity,
independence) if and only if $b_t$ is increasing (decreasing, constant).
\end{theorem}

The intuition behind the proof of this theorem is as follows: Suppose that $%
\delta >0$ (so the modal action is $l$), that the barrier is decreasing, and
that the process stopped at time $t$. The odds that a modal decision is made
in this situation are 
\begin{equation*}
\frac{p^{l}(t)}{p^{r}(t)}=\frac{\mathbb{P}[Z_{t}=b_{t}|\{\tau =t\}\cap
\{|Z_{t}|=b_{t}\}]}{\mathbb{P}[Z_{t}=-b_{t}|\{\tau =t\}\cap
\{|Z_{t}|=b_{t}\}]},
\end{equation*}%
%
%
%
%
%
%
%
%
%
%
%
where $\{\tau =t\}$ is the event that the process $Z$ has not crossed the
barrier before time $t$. From Bayes rule and the formula for the density of
the normal distribution 
\begin{equation*}
\frac{\mathbb{P}[Z_{t}=b_{t}|\{|Z_{t}|=b_{t}\}]}{\mathbb{P}%
[Z_{t}=-b_{t}|\{|Z_{t}|=b_{t}\}]}=\exp \left( \frac{4\delta b_{t}}{\alpha
^{2}}\right),
\end{equation*}
which is a decreasing function of $t$ whenever $b$ is. Moreover, a symmetry
argument using the Brownian bridge shows that the conditioning event $%
\{\tau=t\}$ does not matter.

Theorem \ref{thm:monot} says that the speed-accuracy tradeoff generated by
exogenous stopping can be reversed if stopping is endogenous, i.e., the
stopping time depends on the process $Z_{t}$.

The special case of the constant boundary DDM is well known to arise as the
solution to an optimal sampling problem, for an agent who thinks there are
only two possible states of the world. The next section presents that model
in detail, and then focuses on a related model in which the agent is also
learning about the intensity of her preference.

\section{Optimal Stopping}

\label{sec:opt-stop}

\subsection{Statement of the model}

Both the simple DDM used to explain data from perception tasks and our
uncertain-difference DDM are based on the idea of sequential learning and
optimal stopping. As we will see, the models differ only in their prior, but
this difference leads to substantially different predictions. In the
learning model, the agent doesn't know the true utilities, $\theta =(\theta
^{l},\theta ^{r})\in \mathbb{R}^{2}$, but has a prior belief about them $\mu
_{0}\in \Delta (\mathbb{R}^{2})$. The agent observes a signal $%
(Z_{t}^{i})_{t\in \mathbb{R}_{+}}$ which as in the DDM has the form of a
drift plus a Brownian motion; in the learning model we assume that the drift
of each $Z^{i}$ is equal to the corresponding state, so that 
\begin{equation}
dZ_{t}^{i}=\theta ^{i}dt+\alpha dB_{t}^{i}.
\end{equation}%
where $\alpha $ is the noisiness of the signal and the processes $%
\{B_{t}^{i}\}$ are independent.\footnote{%
This process was also studied by \cite{Natenzon} to study stochastic choice
with exogenously forced stopping times; he allows utilities to be
correlated, which can explain context effects.} The signals and prior lie in
a probability space $(\Omega ,\mathbb{P},\{\mathcal{F}_{t}\}_{t\in \mathbb{R}%
_{+}}),$ where the information $\mathcal{F}_{t}$ that the agent observed up
to time $t$ is simply the paths $\{Z_{s}^{i}\}_{0\leq s<t}.$We denote the
agent's posterior belief about $\theta $ given this information by $\mu _{t}$%
. Let $X_{t}^{i}=\mathbb{E}_{\mu _{t}}\theta ^{i}=\mathbb{E}\left[ \theta
^{i}\mid \mathcal{F}_{t}\right] $ be the posterior mean for each $i=l,r$. As
long as the agent delays the decision she has to pay flow cost, which for
now we assume to be constant $c>0$. (Section \ref{sec:cost} explores the
implications of time varying cost.) The agent's problem is to decide which
option to take and at which time. Waiting longer will lead to more informed
and thus better decisions, but also entails higher costs. What matters for
this decision is the difference between the two utilities, so a sufficient
statistic for the agent is 
\begin{equation*}
Z_{t}:=Z_{t}^{l}-Z_{t}^{r}=(\theta ^{l}-\theta ^{r})t+\alpha \sqrt{2}B_{t},
\end{equation*}%
where $B_{t}=\frac{1}{\sqrt{2}}(B_{t}^{1}-B_{t}^{2})$ is a Brownian Motion.

When the agent stops it is optimal to choose the option with the highest
posterior expected value; thus, the value of stopping at time $t$ is $%
\max_{i=l,r}X_{t}^{i}$. The agent decides optimally when to stop: he chooses
a \emph{stopping time} $\tau $, i.e., a function $\tau :\Omega \rightarrow
\lbrack 0,+\infty ]$ such that $\{\tau \leq t\}\in \mathcal{F}_{t}$ for all $%
t$; let $\mathcal{T}$ be the set of all stopping times. Hence, the problem
of the agent at $t=0$ can be stated as 
\begin{equation}
\max_{\tau \in \mathcal{T}}\mathbb{E}\left[ \max_{i=l,r}X_{\tau }^{i}-c\tau %
\right] \footnote{%
Following the literature, in cases where the optimum is not unique, we focus
on the minimal optimal stopping time.}.  \label{eq:agents_problem}
\end{equation}

With the agent's prior in hand, we can now define a subjective version of
the relationship between speed and accuracy. Note that Definition \ref%
{def:sat} holds in a strong sense: For each possible value of $\theta $ the
agent's choices deteriorate over time. The following definition asks only
that this is true on average, with respect to the subjective probability of
the agent. Let $p^{m}(t,\theta )$\ be the probability of making the modal
choice when the true state is $\theta $ and the agent stops at~$t$.

\begin{definition}
$P$ displays a \textit{subjective speed-accuracy complementarity} iff 
\begin{equation*}
\int\limits_{\theta }p^{m}(t,\,\theta )\mu _{0}(\theta )d\theta
\end{equation*}
is a decreasing function of $t$.
\end{definition}

Thus, under subjective speed-accuracy complementarity, the subjective
probability of the agent of making a correct choice is a decreasing function
of time.

\subsection{Certain Difference}

In the simple DDM the agent's prior is concentrated on two points: $\theta
_{l}=(\theta ^{\prime \prime },\theta ^{\prime })$ and $\theta _{r}=(\theta
^{\prime },\theta ^{\prime \prime })$, where $\theta ^{\prime \prime
}>\theta ^{\prime }.$ The agent receives payoff $\theta ^{\prime \prime }$
for choosing $l$ in state $\theta _{l}$ or $r$ in state $\theta _{r},$ and $%
\theta ^{\prime }<\theta ^{\prime \prime }$ for choosing $r$ in state $%
\theta _{l}$ or $r$ in state $\theta _{l}$, so she knows that the magnitude
of the utility difference between the two choices is $|\theta^{\prime \prime
}-\theta ^{\prime }|$, but doesn't know which action is better. We let $\mu
_{0}$ denote the agent's prior probability of $\theta _{l}.$

This model was first studied in discrete time by \cite{Wald47} (with a
trade-off between type I and type II errors taking the place of utility
maximization) and by \cite{ABG} in a standard dynamic programming setting. A
version of the result for the continuous-time, Brownian-signal case can be
found for example in \cite{Shiryaev69, Shiryaev}.

\begin{theorem}
\label{thm:simpleDDM} With a binomial prior, there is $k>0$ such that the
minimal optimal stopping time is $\hat{\tau}=\inf \{t\geq 0:|l_{t}|=k\}$,
where $l_t=\log \left( \frac{\mathbb{P}\left[ \theta =\theta _{l}\,|\,%
\mathcal{F}_{t}\right] }{\mathbb{P}\left[ \theta =\theta _{r}\,|\,\mathcal{F}%
_{t}\right] }\right)$. Moreover, when $\mu_0 =0.5$, the optimal stopping
time has a DDM representation with a constant boundary $b$: 
\begin{equation*}
\hat{\tau}=\inf \{t\geq 0\colon |Z_{t}|\geq b\}.\footnote{%
This is essentially Theorem 5, p. 185 of \cite{Shiryaev}, but our
formulation is superficially more general, as in his model the drift only
depends on the sign of the utility difference, but not on its magnitude.}
\end{equation*}
\end{theorem}

Theorems \ref{thm:monot} and \ref{thm:simpleDDM} imply that the simple DDM
satisfies speed-accuracy independence and a fortiori subjective speed
accuracy independence. From the point of view of most economic applications,
the simple DDM misses an important feature, as the assumption that the agent
knows the magnitude of the payoff difference rules out cases in which the
agents is learning the intensity of his preference. At the technical level,
the assumption that the utility difference is known, so there are only two
states, implies that the current value of the process $Z_{t}$ is a
sufficient statistic, regardless of the elapsed time; this is why the
stopping boundary $b$ in this model is constant. Intuitively, one might
expect that if $Z_{t}$ is close to zero and $t$ is large, the agent would
infer that he is close to being indifferent between $l $ and $r$ and so
stops, even though for the same value of $Z_{t}$ the agent would choose to
continue when $t$ is small. This inference is ruled out by the binomial
prior, which says that the agent is sure that he is not indifferent. We now
turn to a model with a Gaussian prior which makes such inferences possible.

\subsection{Uncertain-difference DDM}

\label{sec:ud}

In the uncertain-difference DDM, the agent's prior $\mu _{0}$ is independent
for each action and $\mathcal{N}(X_{0},\sigma _{0}^{2})$. Given the
specification of the signal process \eqref{eq:Z}, the posterior $\mu _{t}$
is $\mathcal{N}(X_{t},\sigma _{t}^{2})$, where 
\begin{equation}
X_{t}^{i}=\frac{X_{0}^{i}\sigma _{0}^{-2}+Z_{t}^{i} \alpha^{-2}}{\sigma
_{0}^{-2}+t\alpha^{-2}}\text{\quad and \quad }\sigma _{t}^{2}=\frac{1}{%
\sigma _{0}^{-2}+t\alpha^{-2}}.  \label{eq:X}
\end{equation}

To gain intuition, consider the agent at time $t$ contemplating a strategy
of waiting $dt$ more seconds and stopping then. The agent's utility of
stopping now is $\max_{i=l,r}X_{t}^{i}$. If the agent waits, he will have a
more accurate belief and hence he will be able to make a more informed
decision, but he will pay an additional cost, leading to an expected change
in utility of $\left( \mathbb{E}_{t}\max_{i=l,r}X_{t+dt}^{i}-%
\max_{i=l,r}X_{t}^{i}\right) -c dt$. The main intuition for our results is
that the value of the additional information gained per unit time is
decreasing in $t$, which leads to stopping regions being time-dependent.

The following theorem states this intuition formally.\footnote{%
Time-dependent stopping thresholds also arise if the cost or utility
functions are time-dependent or if there is a fixed terminal date, see e.g. 
\cite{RapoportBurkheimer71} and \cite{Drugowitsch12}.}

\begin{theorem}
\label{thm:qual} Let $\tau^*$ be the minimal optimal strategy in %
\eqref{eq:agents_problem}. Then

\begin{enumerate}
\item There is a strictly decreasing, strictly positive $k:\mathbb{R}%
_{+}\rightarrow \mathbb{R}_{+}$ such that 
\begin{equation*}
\tau ^{\ast }=\inf \{t\geq 0\colon |X_{t}^{l}-X_{t}^{r}|\geq k(t)\}.
\end{equation*}%
Moreover $\lim_{t\rightarrow \infty}k(t)=0$.

\item If $X_{0}^{l}=X_{0}^{r}$, there is a strictly positive $b:\mathbb{R}%
_{+}\rightarrow \mathbb{R}_{+}$ such that 
\begin{equation*}
\tau ^{\ast }=\inf \{t\geq 0\colon |Z_{t}^{l}-Z_{t}^{r}|\geq b_t\}.
\end{equation*}
\end{enumerate}
\end{theorem}

Part (1) of the theorem describes the optimal strategy $\tau ^{\ast }$ in
terms of stopping regions for posterior means $X_{t}^{l}-X_{t}^{r}$: It is
optimal for the agent to stop once the expected utility difference exceeds a
decreasing threshold $k(t)$. This follows from the principle of optimality
for continuous time processes and the shift invariance property of the value
function, which is due to the normality of the posterior. Intuitively, if
the expected utility difference is small for large $t$, the agent concludes
that the two items are most likely indifferent and since $\sigma _{t}^{2}$
is small for large $t$, the expected gain from learning more is low, so the
agent stops. On the other hand, at $t=0$ even if the expected utility
difference is small, $\sigma _{0}^{2}$ is large so the expected value of
learning is high and thus the agent continues.

Part (2) of the theorem describes the optimal strategy $\tau ^{\ast }$ in
terms of stopping regions for the signal process $Z_{t}:=Z_{t}^{l}-Z_{t}^{r}$%
. This facilitates comparisons with the simple DDM, where the process of
beliefs lives in a different space and is not directly comparable.

One way to understand the difference between this model and the one from the
previous section is to consider the agent's posterior beliefs when $%
Z_{t}\approx 0$ for large $t$. In the certain difference model, the agent
interprets the signal as noise, since according to his prior the utilities
of the the two alternatives are a fixed distance apart, so the agent
disregards the signal and essentially starts from anew. This is why the
optimal boundaries are constant in this model. On the other hand, in the
uncertain difference model the agent's interpretation of $Z_{t}\approx 0$
for large $t$ is that the two alternatives are nearly indifferent, which
prompts the agent to stop the costly information gathering process and make
a decision right away. This is why the optimal boundaries are decreasing in
this model.

We do not know of a closed-form solution for the functions $k$ and $b$;
however, we can show that, as functions of the initial variance $\sigma
_{0}^{2}$, $c$, and noisiness $\alpha $, they have to satisfy the following
conditions. The conditions provide useful information about the
identification of the parameters of the model, and about how the predictions
of the model change as the parameters are varied in experiments. They are
also used to show that $k$ is Lipschitz continuous, which simplifies the
analysis of the boundary value problem, and that it declines with time at
rate at least $1/\sqrt{t}$, which is at the heart of the proof of Theorem~%
\ref{thm:monot2} below.

\begin{theorem}
\label{thm:quant} The optimal solution $k(t,c,\sigma _{0},\alpha )$ to
problem \eqref{eq:agents_problem} is Lipschitz continuous in $t$ and
satisfies: 
\begin{align}
k(t,c,\sigma _{0},\alpha )& =k(0,c,\sigma _{t},\alpha )\text{ for all }t\geq
0  \label{eqq:1} \\
k(0,c,\lambda \sigma _{0},\alpha )& =\lambda k(0,c\lambda ^{-3},\sigma
_{0},\alpha )\text{ for all }\lambda >0  \label{eqq:2} \\
k(t,c,\sigma _{0},\lambda \alpha )& =\lambda k(t,\lambda^{-1} c,\lambda
^{-1} \sigma _{0},\alpha )\text{ for all }t, \lambda >0  \label{eqq:4} \\
k(0,\lambda c,\sigma _{0},\alpha )& \geq \lambda ^{-1}k(0,c,\sigma
_{0},\alpha )\text{ for all }\lambda >0.  \label{eqq:3}
\end{align}%
\end{theorem}

The first equality follows from the fact that an agent who starts at time 0
with prior $\mathcal{N}(X_{0},\sigma _{0}^{2})$, has the same beliefs at
each time $t^{\prime }>t$ as an agent who started at time $t$ with prior $%
\mathcal{N}(X_{t},\sigma _{t}^{2}).$ This equality is used at various steps
in the proofs, including showing that $k$ is Lipschitz continuous in $t$,
which is convenient for technical reasons; it allows us to ignore the
complications of viscosity solutions and have an exact solution to the PDE
that characterizes the optimal stopping rule. The proofs of equalities %
\eqref{eqq:2} and \eqref{eqq:4} use a space-time change. Inequality %
\eqref{eqq:3} follows from a space-time change argument and the fact that
more information is always better for the agent.

Using the results of \cite{Bather62} we can characterize the functional form for $\bar{k}$ that
satisfies the above conditions with equality, and the $\bar{b}$ that
corresponds to it.

\begin{fact}
\label{fact:Bather} Let 
\begin{align}
\bar{k}(t,c,\sigma _{0},\alpha )& =\frac{1}{2c\alpha ^{2}(\sigma
_{0}^{-2}+\alpha ^{-2}t)^{2}}  \label{eq:k} \\
\shortintertext{and}
\bar{b}(t,c,\sigma _{0},\alpha )& =\frac{1}{2c\alpha ^{2}(\sigma
_{0}^{-2}+\alpha ^{-2}t)}.  \label{eq:b}
\end{align}
Then $\bar k$ is the only function that satisfies \eqref{eqq:1}--\eqref{eqq:3} with equality, and $\bar b$ is the associated boundary in the signal space. Moreover, there are constants $\beta, T>0$ such that for all $t>T$ 
\begin{equation*}
\left\vert \bar{b}(t,c,\sigma _{0})-b(t,c,\sigma _{0})\right\vert \leq \frac{
\beta }{(\sigma_0^{-2} +\alpha^{-2}t)^{5/2}}.
\end{equation*} 
\end{fact}

Fact \ref{fact:Bather} implies that since $\bar b$ declines to zero at rate $1/t$, this implies that $b$ does
too. Using numerical methods, we have found that $\bar{k}$ and $\bar{b}$ are good
numerical approximations to the optimal boundaries $k$ and $b$.\footnote{%
The approximations obtained by \cite{Bather62} and \cite{vanMoerbeke74} show
that $b\neq \bar{b}$; however our simulations and the bounds from Fact \ref%
{fact:Bather} indicate that $\bar{b} $ approximates the solution quite
accurately, so it may be useful for estimation purposes.} 

Note that if $b$ is decreasing, like $\bar{b}$ is, then by Theorem \ref%
{thm:monot}, any uncertain-difference DDM displays speed-accuracy
complementarity. We have not been able to show this but we can show that the
uncertain-difference DDM has the following weaker property:

\begin{theorem}
\label{thm:monot2} The Gaussian DDM displays subjective speed-accuracy
complementarity
\end{theorem}

This is true because the boundary $k$ is decreasing at the rate at least $1/%
\sqrt{t}$, which, as we show, follows from Theorem \ref{thm:quant}. This
theorem implies that the analyst will observe a speed-accuracy
complementarity in an experiment in which in the agent faces a series of
decisions with states $(\theta ^{r},\theta ^{l})$ and the values of $\theta $
are drawn according to the agent's prior. In particular, as long as the
prior is correct, speed-accuracy complementarity will hold on average; i.e.,
it will hold for the average $P$ in a given experiment. In addition, we
expect that the speed-accuracy complementarity should hold at least
approximately if the agent's beliefs are approximately correct but we have
not shown this formally. Moreover, the complementarity can hold even across
experiments as long as the distributions of the states (both objective and
subjective) are close enough. That is, while we expect choice-accuracy
complementarity to hold within a given class of decision problems, it need
not hold across classes: if $l$ and $r$ are two apartments with a given
utility difference $\delta =\theta ^{l}-\theta ^{r}$, we expect the agent to
spend on average more time here than on a problem where $l$ and $r$ are two
lunch items with the same utility difference $\delta $. This is because we
expect the prior belief of the agent to be domain specific and in
particular, the variance of the prior, $\sigma _{0}^{2}$, to be higher for
houses than for lunch items. Similarly, the complementarity can hold across
subjects (indeed, the data in Figure \ref{fig:1} is a cross-section of
subjects), as long as their boundaries are not too different; in the extreme
case when one subject has a cost much lower than another subject, the first
one will make choices which are longer and more accurate than the choices of
the second subject.

Finally, we note that the uncertain difference model is equivalent to the 
\cite{Chernoff61} ex post regret model, where for any stopping time $\tau $
the objective function is 
\begin{equation*}
\text{\textrm{Ch\thinspace }}(\tau ):=\mathbb{E}\left[ -\mathbf{1}%
_{\{x_{\tau }^{l}\geq x_{\tau }^{r}\}}(\theta ^{r}-\theta ^{l})^{+}-\mathbf{1%
}_{\{x_{\tau }^{r}>x_{\tau }^{l}\}}(\theta ^{l}-\theta ^{r})^{+}-c\tau %
\right] ;
\end{equation*}%
that is, the agent gets zero for making the correct choice and is penalized
the foregone utility for making the wrong choice.

\begin{fact}
\label{fact:Chernoff} For any stopping time $\tau$ 
\begin{equation*}
\text{\textrm{Ch\,}}(\tau)= \mathbb{E}\left[\max\{X^l_\tau, X^r_\tau\}-c\tau %
\right] + \kappa,
\end{equation*}
where $\kappa$ is a constant independent of $\tau$; therefore, these two
objective functions induce the same choice process.
\end{fact}

Chernoff and following authors have focused on the behavior of the optimal
boundary for very small and very large values of $t$. \cite{LaiLim05} say
that Chernoff's heuristic arguments can be adapted to provide a rigorous
proof that the boundary converges to zero, but do not provide details.
Moreover, we have not found any relevant monotonicity results in this
literature.

\subsection{Non-Linear Cost}

\label{sec:cost}

In deriving the DDM representation from optimal stopping, we have so far
have assumed that the cost of continuing per unit time is constant. We have
seen that in the uncertain-difference model, the optimal boundary decreases
due to the fact there is less to learn as time goes on. One would expect
that the boundary could increase if costs decrease sufficiently quickly.
This raises the question of which DDM representations can be derived as a
solution to an optimal stopping problem when the cost is allowed to vary
arbitrarily over time. The next result shows that for any boundary there
exists a cost function such that the boundary is optimal in the learning
problem with normal or binomial priors. Thus optimal stopping on its own
imposes essentially no restrictions on the observed choice process; the
force of the model derives from its joint assumptions about the evolution of
beliefs and the cost function.

\begin{theorem}
\label{thm-implementability} Consider either the Certain or the
Uncertain-Difference DDM. For any finite boundary $b$ and any finite set $%
G\subseteq \mathbb{R}_{+}$ there exists a cost function $d:\mathbb{R}%
_{+}\rightarrow \mathbb{R}$ such that $b$ is optimal in the set of stopping
times $T$ that stop in $G$ with probability one 
\begin{equation*}
\inf \{t\in G\colon |X_{t}|\geq b_{t}\}\in \argmax_{\tau \in \mathcal{T}}%
\mathbb{E}\left[ \max \{X_{\tau }^{1},X_{\tau }^{2}\}-d(\tau )\right] \,.
\end{equation*}
\end{theorem}

In particular, there is a cost function such that the exponentially
decreasing boundaries in \cite{Milosavljevic10} are optimal, and a cost
function such that there is speed-accuracy independence.

Intuitively, the reason this result obtains is that the optimal stopping
rule always takes the form of a cut-off:\ If the agent stops at time $t$
when $X_{t}=x$, she stops at time $t$ whenever $|X_{t}|>x.$ This allows us
to recursively construct a cost function that rationalizes the given
boundary by setting the cost at time $t$ equal the expected future gains
from learning. To avoid some technical issues having to do with the
solutions to PDE's, we consider a discrete-time finite-horizon formulation
of the problem, where the agent is only allowed to stop at times in a finite
set $G$. This lets us construct the associated cost function period by
period instead of using smoothness conditions and stochastic calculus.%
\footnote{%
The proof of the theorem relies on a result on implementable stopping times
from \cite{kruse2014optimal}. In another paper \cite{kruse2014inverse}
generalize this result to continuous time, but as the absolute value is not
covered by their result we can not use it here. Nevertheless, we conjecture
that the methods used in that paper can be extended to prove the result in
continuous time directly.}

\section{Conclusion}

The recent literature in economics and cognitive science uses
drift-diffusion models with time-dependent boundaries. This is helpful in
matching observed properties of reaction times, notably their correlation
with chosen actions, and in particular a phenomenon that we call
speed-accuracy complementarity, where earlier decisions are better than
later ones. In Section \ref{sec:DDM} we showed that the monotonicity
properties of the boundary characterize whether the observed choice process
displays a speed-accuracy complementarity, or the opposite pattern of a
speed-accuracy tradeoff. This ties an observable property of behavior (the
correlation between reaction times and decisions) to an unobservable
construct of the model (the boundary). This connection is helpful for
understanding the qualitative properties of DDMs; it may also serve as a
useful point of departure for future quantitative exploration of the
connection between the rate of decline of the boundary and the strength of
correlation between reaction times and actions.

In Section \ref{sec:opt-stop} we investigated the DDM as a solution to the
optimal sequential sampling problem, where the agent is unsure about the
utility of each action and is learning about it as the time passes,
optimally deciding when to stop. We studied the dependence of the solution
on the nature of the learning problem and also on the cost structure. In
particular, we proposed a model in which the agent is learning not only
about which option is better, but also by how much it is better. We showed
that the boundary in this model declines to zero at the rate $1/t$. We also
showed that any boundary could be optimal if the agent is facing a nonlinear
cost of time.

The analysis of our paper provides a precise foundation for DDMs with
time-varying boundaries and establishes a set of useful connections between
various parameters of the model and predicted behavior, thus enhancing the
theoretical understanding of the model as well as making precise its
empirical content. We expect the forces identified in this paper to be
present in other decisions involving uncertainty: not just in tasks used in
controlled laboratory experiments, but also in decisions involving longer
time scales, such as choosing an apartment rental, or deciding which papers
to publish (as a journal editor). We hope these results will be a helpful
stepping stone for further work.

\newpage

\vspace{2em} \appendix{\noindent {\LARGE \textbf{Appendix: Proofs}}} 
\singlespacing

\section{General Results}

\subsection{{Proof of Fact \protect\ref{fact:FOSD}}}

{\small \noindent \textbf{Proof of Fact \ref{fact:FOSD}}: To prove part (1)
note that by the definition of a conditional distribution 
\citep[property
(c) p. 343 of][]{Dudley} we have $F^i(t)=\int_{[0, t]} \frac{p^i(s)}{P^i}
dF(s)$, so the density of $F^i$ is $f^i(t)=\frac{p^i(t)}{P^i}$. Since $F^m$
is absolutely continuous w.r.t. $F^o$, the ratio $\frac{f^m(t)}{f^o(t)}$ is
well defined $F$-almost everywhere and equals $\frac{p^m(t)}{p^o(t)}\frac{P^m%
}{P^o}$. This expression is increasing (decreasing, constant) if and only if 
$p^m(t)$ is increasing (decreasing, constant). }

{\small To prove part (2), note that by the definition of a conditional
distribution we have 
\begin{equation}
Q^i(t)=\frac{P^i F^i(t)}{F(t)}=\frac{\int_{[0, t]} p^i(s) dF(s)}{F(t)}.
\label{eq:Qp}
\end{equation}
Thus, for $t<t^{\prime }$ we have 
\begin{align*}
Q^l(t) > Q^l(t^{\prime }) &\text{ \ iff\ \ } \frac{\int_{[0, t]} p^l(s)dF(s)%
}{F(t)}\geq \frac{\int_{[0, t]} p^l(s)dF(s) + \int_{(t, t^{\prime }]}
p^l(s)dF(s)}{F(t)+ [F(t^{\prime })-F(t)]} \\
&\text{ \ iff\ \ } \frac{\int_{[0, t]} p^l(s)dF(s)}{F(t)} \geq \frac{%
\int_{(t, t^{\prime }]} p^l(s)dF(s)}{F(t^{\prime })-F(t)},
\end{align*}
which is true if $p^l(\cdot)$ is a decreasing function since the LHS is the
average of $p^l$ on $[0, t]$ and the RHS is the average on $(t, t^{\prime }]$%
. However, the opposite implication may not hold, for example, consider $%
p^l(t):=(t-2/3)^2$ and $F(t)=t$ for $t\in [0, 1]$. Then $p^l(t)$ is not
decreasing, but $Q^l(t)$ is. }

{\small To prove part (3), note that by \eqref{eq:Qp} we have 
\begin{equation*}
F^l(t) > F^r(t) \text{ \ iff\ \ } \frac{Q^l(t)}{P^l}\geq \frac{Q^r(t)}{P^r} 
\text{ \ iff\ \ } Q^l(t)\geq P^l = \lim_{s\to\infty} Q^l(s),
\end{equation*}
where we used the fact that $Q^l(t)+Q^r(t)=1$ and $P^l+P^r=1$. Thus, if $Q^l$
is a decreasing function, the RHS will hold. However, the opposite obviously
doesn't have to hold.\qed
}

\subsection{{Proof of Fact \protect\ref{fact:DDM}}}

{\small Let $P$ be a Borel probability measure on $\Omega:=A\times T$.
Define the process $Z_t$ as follows. For any $\omega=(a^*, t^*)\in \Omega$
let 
\begin{equation*}
Z_t(\omega):=%
\begin{cases}
0 & \text{ if } t<t^* \\ 
+1 & \text{ if } t\geq t^* \text{ and } a^*=l \\ 
-1 & \text{ if } t\geq t^* \text{ and } a^*=r.%
\end{cases}%
\end{equation*}
Note that $Z_t$ is Borel measurable for each $t$, so $Z_t$ is a stochastic
process on the probability space $(\Omega, \mathcal{B}_{\Omega}, P)$. Set $%
b:=1$ and observe that $(Z_t, b)$ is a DDM-like representation of $P$. }

{\small Note that a representation with continuous paths is possible, where 
\begin{equation*}
Z_t(\omega):=%
\begin{cases}
\frac{t}{t^*} & \text{ if } a^*=l \\ 
-\frac{t}{t^*} & \text{ if } a^*=r;%
\end{cases}%
\end{equation*}
however, in this representation the filtration generated by $Z_t$ is not
right-continuous at zero. \qed
}

\subsection{{Proof of Theorem \protect\ref{thm:monot}}}

{\small Let $f:R_{+}\rightarrow R_{+}$ be the density of the distribution of
stopping times, and $g:R_{+}\times R\rightarrow R_{+}$ be the density of $%
Z_{t}$ i.e 
\begin{equation*}
g(t,y)=\frac{\partial }{\partial y}\mathbb{P}\left[ Z_{t}\leq y|\delta
,\alpha \right] =\frac{\partial }{\partial y}\mathbb{P}\left[ \frac{B_{t}}{%
\sqrt{t}}\leq \frac{y-\delta t}{\alpha \sqrt{t}}\right] =\phi \left( \frac{%
y-\delta t}{\alpha \sqrt{t}}\right) ,
\end{equation*}%
where $\phi (x)=\frac{1}{\sqrt{2\pi }}e^{-x^{r}/2}$ is the density of the
standard normal. By Bayes rule: 
\begin{align*}
p^{l}(t)& =\mathbb{P}[Z_{t}=b_{t}|\tau =t,\delta ,\alpha ]=\frac{g(t,b_{t})%
\mathbb{P}[\tau =t|Z_{t}=b_{t},\delta ,\alpha ]}{f(t)} \\
p^{r}(t)& =\mathbb{P}[Z_{t}=-b_{t}|\tau =t,\delta ,\alpha ]=\frac{g(t,-b_{t})%
\mathbb{P}[\tau =t|Z_{t}=-b_{t},\delta ,\alpha ]}{f(t)}
\end{align*}%
} {\small It follows from $Z_{0}=0$ and the symmetry of the upper and the
lower barrier that 
\begin{equation}
\mathbb{P}[\tau =t|Z_{t}=b_{t},\delta ,\alpha ]=\mathbb{P}[\tau
=t|Z_{t}=-b_{t},-\delta ,\alpha ],  \label{eq:symmetry}
\end{equation}%
because for any path of $Z$ that ends at $b_{t}$ and crosses any boundary
before $t$, the reflection of this path ends at $-b_{t}$ and crosses some
boundary at the same time. }

{\small The induced probability measure over paths conditional on $Z_t=b_t$
is the same as the probability of the Brownian Bridge.\footnote{%
See, e.g., Proposition 12.3.2 of \cite{Dudley} or Exercise 3.16, p. 41 of 
\cite{RY}.} The Brownian Bridge is the solution to the SDE $dZ_{s}=-\frac{%
b_t-Z}{t-s}ds+\alpha dB_{s}$ and notably does not depend on the drift $%
\delta $, which implies that 
\begin{equation}
\mathbb{P}[\tau=t | Z_t=-b_t, -\delta, \alpha]=\mathbb{P}[\tau=t | Z_t=-b_t,
\delta, \alpha]  \label{eq:drift-invariance}
\end{equation}
} {\small Thus, by \eqref{eq:symmetry} and \eqref{eq:drift-invariance} we
have that 
\begin{equation*}
\frac{p^{l}(t)}{p^{r}(t)}=\frac{g(t,b_t)}{g(t,-b_t)}=\exp \left( \frac{%
4\delta b_t}{\alpha ^{2}}\right) .
\end{equation*}%
Wlog $m=l$ and $\delta >0$; the above expression is decreasing over time if
and only if $b_t$ is decreasing over time. \qed}

\section{The Uncertain-Difference Model}

\subsection{The Value Function}

{\small Our results use on the following representation of the posterior
process in the uncertain-difference model. }

\begin{lemma}
{\small \label{lem:repr} For any $t>0$ 
\begin{equation*}
X_t^i=X_0^i + \int_0^t \frac{\alpha^{-1}}{\sigma_{0}^{-2}+s\alpha^{-2}}
dW_{s}^{i}
\end{equation*}
where $W_{s}^i$ is a Brownian motion with respect to the filtration
information of the agent. }
\end{lemma}

{\small \noindent \textbf{Proof}: This follows from Theorem 10.1 and
equation 10.52 of \cite{LS} by setting $a=b=0$ and $A=1, B=\alpha$. \qed
}

{\small \medskip Define the continuation value as the expected value an
agent can achieve by using the optimal continuation strategy if he beliefs
the posterior means to be $(x^{l},x^{r})$ at time $t$ and the variance of
his prior equaled $\sigma _{0}^{2}$ at time 0 and the noisiness of the
signal is $\alpha$. 
\begin{equation*}
V(t,x^{l},x^{r},c,\sigma _{0}, \alpha):=\sup_{\tau \geq t}\mathbb{E}_{(t,
x^{l},x^{r},\sigma _{0},\alpha)}\left[ \max \{X_{\tau }^{l},X_{\tau
}^{r}\}-c\,(\tau -t)\right] \,.
\end{equation*}
}

\begin{lemma}
{\small \label{lem:properties-continuation-value} The continuation value has
the following properties: }

\begin{enumerate}
\item {\small $\mathbb{E}_{(t, x^{l},x^{r},\sigma
_{0},\alpha)}\max\{\theta^l,\theta^r\}\geq V(t,x^l,x^r,c,\sigma_{0},
\alpha)\geq\max\{x^l,x^r\}.$ }

\item {\small $V(t,x^l,x^r,c,\sigma_{0}, \alpha)-k=V(t,x^l-k,x^r-k,
c,\sigma_{0}, \alpha)$ for every $k\in\mathbb{R}.$ }

\item {\small The option value $V(t,x^l,x^r,c,\sigma_{0}, \alpha)-x^{i}$ is
decreasing in $x^{i}$ for $i\in\{l,r\}$. }

\item {\small $V(t,x^l,x^r,c,\sigma_{0}, \alpha)$ is increasing in $x^l$ and 
$x^r.$ }

\item {\small The continuation value is Lipschitz continuous in $x^l$ and $%
x^r$. }
\end{enumerate}
\end{lemma}

\subsection*{{{Proof of Lemma \protect\ref{lem:properties-continuation-value}%
}}}

{\small In this proof we equivalently represent a continuation strategy by a
pair of stopping times $(\tau^l,\tau^r)$, one for each alternative. }

{\small \noindent \textbf{Proof of 1}: For the lower bound, the agent can
always stop immediately and get $x^l$ or $x^r$. For the upper bound, the
agent can't do better than receiving a fully informative signal right away
and pick the better item immediately. }

{\small \noindent \textbf{Proof of 2}: Fix a continuation strategy $%
(\tau^l,\tau^r)$; the expected payoff equals 
{\footnotesize
\begin{align*}
& \mathbb{E}\left[\mathrm{\mathbf{1}}_{\{\tau^l\leq\tau^r\}}X_{\tau^l}^l+%
\mathrm{\mathbf{1}}_{\{\tau^l>\tau^r\}}X_{\tau^r}^r-c\left(\min\{\tau^l,%
\tau^r\}-t\right)\,|\, X_{t}^l=x^l-k,X_{t}^r=x^r-k\right] \\
& = \mathbb{E}\bigg[\mathrm{\mathbf{1}}_{\{\tau^l\leq\tau^r\}}\left(%
\int_{t}^{\tau^l}\frac{\alpha^{-1}}{\sigma_0^{-2}+s\alpha^{-2}}\mathrm{d}%
W_{s}^l+x^l-k\right)+\mathrm{\mathbf{1}}_{\{\tau^l>\tau^r\}}\left(\int_{t}^{%
\tau^r}\frac{\alpha^{-1}}{\sigma_0^{-2}+s\alpha^{-2}}\mathrm{d}%
W_{s}^r+x^r-k\right) \\
& \qquad \qquad \qquad \qquad \qquad \qquad \qquad \qquad \qquad \qquad
-c\left(\min\{\tau^l,\tau^r\}-t\right)\,|\,X_{t}^l=x^l,X_{t}^r=x^r\bigg] \\
& = \mathbb{E}\left[\mathrm{\mathbf{1}}_{\{\tau^l\leq\tau^r\}}X_{\tau^l}^l+%
\mathrm{\mathbf{1}}_{\{\tau^l>\tau^r\}}X_{\tau^r}^r-c\left(\min\{\tau^l,%
\tau^r\}-t\right)\,|\, X_{t}^l=x^l,X_{t}^r=x^r\right]-k\,.
\end{align*}}
Intuitively, this comes from the translation invariance of the Brownian
motion, i.e., the distribution of $X^i_s$ conditional on $X^i_t=x^i-k$ is
the same as the distribution of $X^i_s-k$ conditional on $X^i_t=x^i$. As $V$
is defined as the supremum over all continuation strategies $(\tau^l,\tau^r)$
the result follows. \\}

{\small \noindent \textbf{Proof of 3}: The expected difference between
stopping at time $t$ with option $l$ and using the continuation strategy $%
(\tau^l,\tau^r) $ is 
{\footnotesize
\begin{align*}
& \mathbb{E}\left[\mathrm{\mathbf{1}}_{\{\tau^l\leq\tau^r\}}X_{\tau^l}^l+%
\mathrm{\mathbf{1}}_{\{\tau^l>\tau^r\}}X_{\tau^r}^r-c\left(\min\{\tau^l,%
\tau^r\}-t\right)\,|\, X_{t}^l=x^l,X_{t}^r=x^r\right]-x^l \\
= & \mathbb{E}\left[\mathrm{\mathbf{1}}_{\{\tau^l\leq\tau^r\}}(X_{%
\tau^l}^l-x^l)+\mathbf{1}_{\{\tau^l>\tau^r\}}(X_{\tau^r}^r-x^l)-c\left(\min%
\{\tau^l,\tau^r\}-t\right)\,|\, X_{t}^l=x^l,X_{t}^r=x^r\right] \\
= & \mathbb{E}\left[\mathrm{\mathbf{1}}_{\{\tau^l\leq\tau^r\}}\int_{t}^{%
\tau^l}\frac{\alpha^{-1}}{\sigma_0^{-2}+s\alpha^{-2}}\mathrm{d}W_{s}^l+%
\mathrm{\mathbf{1}}_{\{\tau^l>\tau^r\}}(X_{\tau^r}^r-x^l)-c\left(\min\{%
\tau^l,\tau^r\}-t\right)\,|\, X_{t}^l=x^l,X_{t}^r=x^r\right]
\end{align*}}
Note that the first part is independent of $x^l$, and $(X_{\tau^r}^r-x^l)$
is weakly decreasing in $x^l$. As for every fixed strategy $(\tau^l,\tau^r)$
the value of waiting is decreasing the supremum over all continuation
strategies is also weakly decreasing in $x^l.$ Thus it follows that the
difference between continuation value $V(t,x^l,x^r,c,\sigma_{0}, \alpha)$
and value of stopping immediately on the first arm $x^l$ is decreasing in $%
x^l$ for every $t$ and every $x^r$. \\}

{\small \noindent \textbf{Proof of 4}: The expected value of using the
continuation strategy $(\tau^l,\tau^r)$ equals 
{\footnotesize
\begin{align*}
& \mathbb{E}\left[\mathrm{\mathbf{1}}_{\{\tau^l\leq\tau^r\}}X_{\tau^l}^l+%
\mathrm{\mathbf{1}}_{\{\tau^l>\tau^r\}}X_{\tau^r}^r-c\left(\min\{\tau^l,%
\tau^r\}-t\right)\,|\, X_{t}^l=x^l,X_{t}^r=x^r\right] \\
= & \mathbb{E}\left[\mathrm{\mathbf{1}}_{\{\tau^l\leq\tau^r\}}\left(%
\int_{t}^{\tau^l}\frac{\alpha^{-1}}{\sigma_0^{-2}+s\alpha^{-2}}\mathrm{d}%
W_{s}^l+x^l\right)+\mathrm{\mathbf{1}}_{\{\tau^l>\tau^r\}}X_{\tau^r}^r-c%
\left(\min\{\tau^l,\tau^r\}-t\right)\,|\, X_{t}^l=x^l,X_{t}^r=x^r\right] \\
= & \mathbb{E}\left[\mathrm{\mathbf{1}}_{\{\tau^l\leq\tau^r\}}\int_{t}^{%
\tau^l}\frac{\alpha^{-1}}{\sigma_0^{-2}+s\alpha^{-2}}\mathrm{d}W_{s}^l+%
\mathrm{\mathbf{1}}_{\{\tau^l>\tau^r\}}X_{\tau^r}^r-c\left(\min\{\tau^l,%
\tau^r\}-t\right)\,|\, X_{t}^l=x^l,X_{t}^r=x^r\right] \\
& +x^l\mathbb{E}\left[\mathrm{\mathbf{1}}_{\{\tau^l\leq\tau^r\}}\,|\,
X_{t}^l=x^l,X_{t}^r=x^r\right],
\end{align*}}
which is weakly increasing in $x^l$. Consequently, the supremum over all
continuation strategies $(\tau^l,\tau^r)$ is weakly increasing in $x^l$. By
the same argument it follows that $V(t,x^l,x^r,c,\sigma_{0}, \alpha)$ is
increasing in $x^r.$\\ }

{\small \noindent \textbf{Proof of 5}: The following argument establishes
that the value function is Lipschitz continuous with constant one. The
initial belief about the mean of either option is additively separable from
the change in belief caused by the information the agent observed after time
zero. Thus changing the initial beliefs moves the posterior beliefs
linearly. Hence, for any fixed stopping time the change in initial belief
can at most linearly move the posterior beliefs about the mean. Furthermore,
the expected cost are unaffected by a change in prior beliefs. Thus, the
supremum over all stopping times can at most be linearly affected by a
change in initial belief. To see this explicitly, observe that 
{\footnotesize 
\begin{align*}
& |V(0,x^l,x^r,c,\sigma_{0}, \alpha)-V(0,y^l,x^r,c,\sigma_{0}, \alpha)| \\
& = \bigg|\sup_{\tau}\mathbb{E}\left[\max\{x^l+\int_{0}^{\tau}\frac{%
\alpha^{-1}}{\sigma_{0}^{-2}+s\alpha^{-2}}\mathrm{d}W_{s}^{1},x^r+\int_{0}^{%
\tau}\frac{\alpha^{-1}}{\sigma_{0}^{-2}+s\alpha^{-2}}\mathrm{d}%
W_{s}^{2}\}-c(\tau)\right] \\
& \qquad\qquad\qquad-\sup_{\tau}\mathbb{E}\left[\max\{y^l+\int_{0}^{\tau}%
\frac{\alpha^{-1}}{\sigma_{0}^{-2}+s\alpha^{-2}}\mathrm{d}%
W_{s}^{1},x^r+\int_{0}^{\tau}\frac{\alpha^{-1}}{\sigma_{0}^{-2}+s\alpha^{-2}}%
\mathrm{d}W_{s}^{2}\}-c(\tau)\right]\,\bigg| \\
& \leq \bigg|\sup_{\tau}\mathbb{E}\bigg[\max\{x^l+\int_{0}^{\tau}\frac{%
\alpha^{-1}}{\sigma_{0}^{-2}+s\alpha^{-2}}\mathrm{d}W_{s}^{1},x^r+\int_{0}^{%
\tau}\frac{\alpha^{-1}}{\sigma_{0}^{-2}+s\alpha^{-2}}\mathrm{d}W_{s}^{2}\} \\
& \qquad\qquad \qquad -\max\{y^l+\int_{0}^{\tau}\frac{\alpha^{-1}}{%
\sigma_{0}^{-2}+s\alpha^{-2}}\mathrm{d}W_{s}^{1},x^r+\int_{0}^{\tau}\frac{%
\alpha^{-1}}{\sigma_{0}^{-2}+s\alpha^{-2}}\mathrm{d}W_{s}^{2}\}\bigg]\bigg| %
\leq |y^l-x^l|\,.\qed
\end{align*}
} }

\subsection{{\ {Proof of Theorem \protect\ref{thm:qual}}}}

\subsubsection{ Characterization}

{\small Note that due to the symmetry of the problem $V(t,x^{l},x^{r},c,%
\sigma _{0}, \alpha)=V(t,x^{r},x^{l},c,\sigma _{0}, \alpha).$ Without loss
of generality suppose $x^{l}\leq x^{r}$ otherwise swap $x^{l}$ and $x^{r}$.
As $X_{t}$ is a Markov process, the principle of optimality\footnote{%
Our model does not satisfy condition (2.1.1) of \cite{PeskirShiryaev}
because for some stopping times the expected payoff is minus infinity, but
as they indicate on p. 2 the proof can be extended to our case.} implies
that the agent's problem admits a solution of the form $\tau =\inf \{t\geq
0:\max_{i=l,r}X_{t}^{i}\geq V(t,X_{t}^{l},X_{t}^{r},c,\sigma _{0}, \alpha)\}$%
. Thus, it is optimal to stop if and only if 
\begin{equation*}
0=V(t,x^{l},x^{r},c,\sigma _{0}, \alpha)-\max
\{x^{l},x^{r}\}=V(t,x^{l},x^{r},c,\sigma _{0},
\alpha)-x^{r}=V(t,x^{l}-x^{r},0,c,\sigma _{0}, \alpha)\,.
\end{equation*}%
An envelope argument yields that $V$ is continuous. Using the continuity and
the monotonicity of $V$ we can define the function $k$ implicitly by 
\begin{equation*}
k(t):=\inf \{x\in \mathbb{R}\colon 0=V(t,-x,0,c,\sigma _{0}, \alpha)\}.
\end{equation*}%
As $x^{l}-x^{r}\leq 0$, $V$ is monotone increasing in the second argument
and $V(t,x^{l}-x^{r},0,c,\sigma _{0}, \alpha)\geq 0$ by Lemma \ref%
{lem:properties-continuation-value} we have 
\begin{equation*}
\{0=V(t,x^{l}-x^{r},0,c,\sigma _{0})\}=\{x^{l}-x^{r}\leq
-k(t)\}=\{|x^{l}-x^{r}|\geq k(t)\}\,.
\end{equation*}%
Hence the optimal strategy equals $\tau =\inf \{t\geq
0:|X_{t}^{l}-X_{t}^{r}|\geq k(t)\}\,.$ }

\subsubsection{Monotonicity}

{\small First, we show that $V(t,x^{l},x^{r},c,\sigma _{0},\alpha )$ is
decreasing in $t$. Note that by Doob's optional sampling theorem for every
fixed stopping strategy $\tau $ 
\begin{eqnarray*}
\mathbb{E}\left[ \max \{X_{\tau }^{l},X_{\tau }^{r}\}-c\tau \mid
X_{t}=(x^{l},x^{r})\right] &=&\mathbb{E}\left[ \max \{X_{\tau }^{l}-X_{\tau
}^{r},0\}+X_{\tau }^{r}-c\tau \mid X_{t}=(x^{l},x^{r})\right] \\
&=&\mathbb{E}\left[ \max \{X_{\tau }^{l}-X_{\tau }^{r},0\}-c\tau \mid
X_{t}=(x^{l},x^{r})\right] +X_{t}^{r}\,.
\end{eqnarray*}%
Define the process $X_{t}:=X_{t}^{l}-X_{t}^{r}$, and note that 
{\footnotesize 
\begin{equation}
X_{t}=X_{t}^{l}-X_{t}^{r}=X_{0}^{l}-X_{0}^{r}+\int_{0}^{t}\frac{\alpha ^{-1}%
}{\sigma _{0}^{-2}+s\alpha ^{-2}}\left( \mathrm{d}W_{s}^{l}-\mathrm{d}%
W_{s}^{r}\right) =X_{0}^{l}-X_{0}^{r}+\int_{0}^{t}\frac{\sqrt{2}\alpha ^{-1}%
}{\sigma _{0}^{-2}+s\alpha ^{-2}}\mathrm{d}\tilde{W}_{s}\,,
\label{eq:Xsqrt2}
\end{equation}%
where $\tilde{W}$ is a Brownian motion. Define a time change as follows: Let 
$q(k)$ solve 
$k=\int_{0}^{q(k)}\left( \frac{\sqrt{2}\alpha ^{-1}}{\sigma
_{0}^{-2}+s\alpha ^{-2}}\right) ^{2}\mathrm{d}s.$ 
This implies that $q(k)=\frac{k\alpha ^{2}\sigma _{0}^{-2}}{2\sigma
_{0}^{2}-k}$. Define $\psi (t)=\frac{2\sigma _{0}^{2}t}{\alpha ^{2}\sigma
_{0}^{-2}+t}$. B}y the Dambis, Dubins--Schwarz theorem 
\citep[see, e.g.,
Theorem 1.6, chapter V of][]{RY} $W_{s}:=(X_{q(s)})_{s\in \lbrack 0,2\sigma
_{0}^{-2}]}$ is a Brownian motion and thus we can rewrite the problem as 
{\footnotesize 
\begin{eqnarray*}
V(t,x^{l},x^{r},c,\sigma _{0},\alpha ) &=&\sup_{\tau \geq \psi (t)}\mathbb{E}%
\left[ \max \{W_{\tau },0\}-c\bigg(q(\tau )-q(\psi (t))\bigg)\,|\,W_{\psi
(t)}=x^{l}-x^{r}\right] +x^{r} \\
&=&\sup_{\tau \geq \psi (t)}\mathbb{E}\left[ \max \{W_{\tau },0\}-c\bigg(%
\int_{\psi (t)}^{\tau }q^{\prime }(s)\mathrm{d}s\bigg)\,|\,W_{\psi
(t)}=x^{l}-x^{r}\right] +x^{r}, \\
&=&\sup_{\tau \geq \psi (t)}\mathbb{E}\left[ \max \{W_{\tau },0\}-c\bigg(%
\int_{\psi (t)}^{\tau }\frac{2\alpha ^{2}}{(2\sigma _{0}^{2}-s)^{2}}\mathrm{d%
}s\bigg)\,|\,W_{\psi (t)}=x^{l}-x^{r}\right] +x^{r}.
\end{eqnarray*}%
} }

{\small Next, we remove the conditional expectation in the Brownian motion
by adding the initial value 
\begin{equation*}
V(t,x^l,x^r,c,\sigma_{0}, \alpha)=\sup_{\tau \geq \psi (t)}\mathbb{E}\left[
\max \{W_{\tau }+\left( x^l-x^r\right) ,0\}-c\int_{\psi (t)}^{\tau }\frac{%
2\alpha ^{2}}{(2\sigma_0^2-s)^2}\mathrm{d}s\right] +x^r\,.
\end{equation*}%
Define $\hat{\tau}=\tau -\psi $ and let wlog $x^l<x^r$, then  
\begin{equation*}
V(t,x^l,x^r,c,\sigma_{0}, \alpha)=\sup_{\hat{\tau}\geq 0}\mathbb{E}\left[
\max \{W_{\tau }-|x^l-x^r|,0\}-c\int_{\psi (t)}^{\psi (t)+\hat{\tau}}\frac{%
2\alpha ^{2}}{(2\sigma_0 ^2-s)^2}\mathrm{d}s\right] +\max \{x^l,x^r\};
\label{eq:ValueFunctionBrownian}
\end{equation*}
because the current state is a sufficient statistic for Brownian motion we
have 
\begin{equation*}
V(t,x^l,x^r,c,\sigma_{0}, \alpha)=\sup_{\hat{\tau}\geq 0}\mathbb{E}\left[
\max \{W_{\tau }-|x^l-x^r|,0\}-c\int_{0}^{\hat{\tau}}\frac{2\alpha ^{2}}{%
(2\sigma_0 ^2-s-\psi (t))^2}\mathrm{d}s\right] +\max \{x^l,x^r\}.
\end{equation*}
Note that for every fixed strategy $\tau $ the cost term is increasing in $t$
and $\psi (t)$ and thus $V(t,x^l,x^r,c,\sigma_{0}, \alpha)-\max \{x^l,x^r\}$
is non-increasing. }

{\small Finally, suppose that $t<t^{\prime }$; then 
\begin{equation*}
0=V(t,-k(t, c, \sigma_0, \alpha),0,c,\sigma_{0}, \alpha)\geq V(t^{\prime
},-k(t, c, \sigma_0, \alpha),0,c,\sigma_{0}, \alpha)\,.
\end{equation*}
By Lemma \ref{lem:properties-continuation-value}, $V(t^{\prime },-k(t, c,
\sigma_0, \alpha),0,c,\sigma_{0}, \alpha)\geq0 $ and hence $0=V(t^{\prime
},-k(t, c, \sigma_0, \alpha),0,c,\sigma_{0}, \alpha)$. Hence 
\begin{equation*}
k(t, c, \sigma_0, \alpha)\geq\inf\{x\in\mathbb{R}\colon0=V(t^{\prime
},-x,0,c,\sigma_{0}, \alpha)\}=k(t^{\prime }, c, \sigma_0, \alpha)\,.
\end{equation*}
}

\subsubsection{Positivity}

{\small The payoff of the optimal decision rule is at least as high as the
payoff from using the strategy that stops at time $\Delta $ for sure.
Because the information gained over a short time period $\Delta $ is of
order $\epsilon ^{\frac{1}{2}}$ and the cost is linear, we expect that it is
always worth buying some information when the expected utility of both
options is the same. To see this formally, note that {\footnotesize 
\begin{eqnarray*}
V(t,x,x,c,\sigma _{0},\alpha )-x &=&\sup_{\tau }\mathbb{E}\left[ \max
\{W_{\tau },0\}-\int_{0}^{\tau }\frac{2c\alpha ^{2}}{(2\sigma
_{0}^{2}-s-\psi (t))^{2}}\mathrm{d}s\right] \\
&\geq &\mathbb{E}\left[ \max \{W_{\epsilon },0\}-\int_{0}^{\epsilon }\frac{%
2c\alpha ^{2}}{(2\sigma _{0}^{2}-s-\psi (t))^{2}}\mathrm{d}s\right] \\
&=&\int_{0}^{\infty }z\frac{1}{\sqrt{2\pi \epsilon }}e^{-\frac{z^{2}}{%
2\epsilon }}\mathrm{d}z-\int_{0}^{\epsilon }\frac{2c\alpha ^{2}}{(2\sigma
_{0}^{2}-s-\psi (t))^{2}}\mathrm{d}s \\
&\geq &\sqrt{\frac{\epsilon }{2\pi }}-\frac{2c\alpha ^{2}\epsilon }{(2\sigma
_{0}^{2}-\psi (t)-\epsilon )^{2}} \\
&\geq &\sqrt{\frac{\epsilon }{2\pi }}-\frac{2c\alpha ^{2}\epsilon }{(2\sigma
_{0}^{2}-\psi (t)-\tilde{\epsilon})^{2}}\text{ for all fixed }\tilde{\epsilon%
}\in \lbrack \epsilon ,2\sigma _{0}^{2}-\psi (t)\,)\,.
\end{eqnarray*}%
} As the first term goes to zero with the speed of square root while the
second term shrinks linearly we get that $V(t,x,x,c,\sigma _{0},\alpha
)-\max \{x,x\}>0$ for some small $\epsilon >0$ and thus the agent does not
stop when his posterior mean is the same on both options. }

\subsubsection{ Zero limit}

{\small Let $k(s, c, \sigma_0, \alpha)\geq K^{*}>0$ for all $s\geq t$.
Consider the time $t$ history where $X_{t}^l=X_{t}^r$. The probability that
the agent never stops (and thus pays infinity costs) is bounded from below
by the probability that the process $X^l-X^r$ stays in the interval $%
[-K^{*},K^{*}]$, 
\begin{equation*}
\mathbb{P}\left[ \sup_{s\in \lbrack t,\infty )}|X_{s}^{l}-X_{s}^{r}|\leq
K(s)|X_{t}^{l}=X_{t}^{r}\right] \geq \mathbb{P}\left[ \sup_{s\in \lbrack
t,\infty )}|X_{s}^{l}-X_{s}^{r}|\leq K^{\ast }|X_{t}^{l}=X_{t}^{r}\right]
\end{equation*}
By the above time change argument this equals the probability that a
Brownian motion leaves the interval $[-K,K]$ in the time from $\psi(t)$ to $%
2\sigma_0^2$, 
\begin{equation*}
\mathbb{P}\left[ \sup_{s\in \lbrack t,\infty )}|X_{s}^{l}-X_{s}^{r}|\leq
K^{\ast }|X_{t}^{l}=X_{t}^{r}\right] =\mathbb{P}\left[ \sup_{s\in \lbrack
\psi (t),2\sigma_0 ^{2}]}|W_{s}|\leq K^{\ast }(s)\right]
\end{equation*}
This probability is non-zero. Thus, there is a positive probability the
agent incurs infinite cost. Because the expected gain is bounded by the full
information payoff, this is a contradiction. \qed }

\subsection{{{Proof of Theorem \protect\ref{thm:quant}}}}

\begin{lemma}
\label{lem:time-change} {\footnotesize \ 
\begin{align*}
V(0,& x^{l},x^{r}, c\lambda ,\sigma _{0},\alpha )= \\
&=\lambda ^{-1}\sup_{\tau ^{\prime }}\mathbb{E}\left[ \max \left\{ \lambda
x^{l}+\int_{0}^{\tau ^{\prime }}\frac{\alpha ^{-1}}{\sigma
_{0}^{-2}+s\lambda ^{-2}\alpha ^{-2}}\mathrm{d}M_{s}^{l},\lambda
x^{r}+\int_{0}^{\tau ^{\prime }}\frac{\alpha ^{-1}}{\sigma
_{0}^{-2}+s\lambda ^{-2}\alpha ^{-2}}\mathrm{d}M_{s}^{r}\right\} -c\tau %
\right]
\end{align*}%
}
\end{lemma}

\noindent \textbf{Proof}: {\small We have that $V(0,x^{l},x^{r},c\lambda
,\sigma _{0},\alpha )$ equals {\footnotesize 
\begin{align*}
& \sup_{\tau }\mathbb{E}\left[ \max \left\{ x^{l}+\int_{0}^{\tau }\frac{%
\alpha ^{-1}}{\sigma _{0}^{-2}+s\alpha ^{-2}}\mathrm{d}W_{s}^{l},x^{r}+%
\int_{0}^{\tau }\frac{\alpha ^{-1}}{\sigma _{0}^{-2}+s\alpha ^{-2}}\mathrm{d}%
W_{s}^{r}\right\} -c\lambda \,\tau \right] \\
& =\lambda ^{-1}\sup_{\tau }\mathbb{E}\left[ \max \left\{ \lambda
x^{l}+\int_{0}^{\tau }\frac{\lambda \alpha ^{-1}}{\sigma _{0}^{-2}+s\alpha
^{-2}}\mathrm{d}W_{(s\lambda ^{2})\lambda ^{-2}}^{l},\lambda
x^{r}+\int_{0}^{\tau }\frac{\lambda \alpha ^{-1}}{\sigma _{0}^{-2}+s\alpha
^{-2}}\mathrm{d}W_{(s\lambda ^{2})\lambda ^{-2}}^{r}\right\} -c\lambda
^{2}\,\tau \right] \\
& =\lambda ^{-1}\sup_{\tau }\mathbb{E}\left[ \max \left\{ \lambda
x^{l}+\int_{0}^{\tau \lambda ^{2}}\frac{\lambda \alpha ^{-1}}{\sigma
_{0}^{-2}+s\lambda ^{-2}\alpha ^{-2}}\mathrm{d}W_{s\lambda
^{-2}}^{l},\lambda x^{r}+\int_{0}^{\tau \lambda ^{2}}\frac{\lambda \alpha
^{-1}}{\sigma _{0}^{-2}+s\lambda ^{-2}\alpha ^{-2}}\mathrm{d}W_{s\lambda
^{-2}}^{r}\right\} -c\lambda ^{2}\,\tau \right] \\
& =\lambda ^{-1}\sup_{\tau }\mathbb{E}\left[ \max \left\{ \lambda
x^{l}+\int_{0}^{\tau \lambda ^{2}}\frac{\alpha ^{-1}}{\sigma
_{0}^{-2}+s\lambda ^{-2}\alpha ^{-2}}\mathrm{d}M_{s}^{l},\lambda
x^{r}+\int_{0}^{\tau \lambda ^{2}}\frac{\alpha ^{-1}}{\sigma
_{0}^{-2}+s\lambda ^{-2}\alpha ^{-2}}\mathrm{d}M_{s}^{r}\right\} -c\lambda
^{2}\,\tau \right]
\end{align*}%
For the step from the second to third line apply Proposition 1.4 of Chapter
V of \cite{RY} with $C_{s}:=s\lambda ^{-2}$and $H_{s}:=\frac{\alpha
^{-1}\lambda }{\sigma _{0}^{-2}+\alpha ^{-2}\lambda ^{-2}s}$ (pathwise to
the integrals with limits $\tau $ and $\tau \lambda ^{2})$. In the next step
we apply a time-change, where $M_{r}^{i}:=\lambda W_{r\lambda ^{-2}}^{i}$ is
a Brownian motion and $\tau ^{\prime}$ is a stopping time measurable in the
natural filtration generated by $M$. \qed} }

\subsubsection{{{{Proof of \eqref{eqq:1}}}}}

{\small Note that Lemma \ref{lem:repr} implies that for any $t<t^{\prime }$ 
\begin{equation*}
X_{t^{\prime }}^{i}=X_{t}^{i}+\int_{0}^{t^{\prime }-t}\frac{\alpha ^{-1}}{%
(\sigma _{0}^{-2}+\alpha ^{-2}t)+\alpha ^{-2}s}dW_{t+s}^{i}
\end{equation*}%
where $W_{s}^{i}$ is a Brownian motion with respect to the filtration
information of the agent. Thus, if the agent starts with a prior at time 0
equal to $\mathcal{N}(X_{0},\sigma _{0}^{2})$, then his belief at time $%
t^{\prime }$ is exactly the same as if he started with a prior at $t$ equal
to $\mathcal{N}(X_{t},\sigma _{t}^{2})$ where $\sigma _{t}^{-2}=\sigma
_{0}^{-2}+\alpha ^{-2}t$. Thus, $V(t,x^{l},x^{r},c,\sigma _{0},\alpha )
=V(0,x^{l},x^{r},c,\sigma _{t},\alpha )$ so 
\begin{align*}
k(t,c,\sigma _{0},\alpha )& =\inf \{x>0\colon 0=V(t,0,-x,c,\sigma
_{0},\alpha )\} \\
&=\inf \{x>0\colon 0=V(0,0,-x,c,\sigma _{t},\alpha )\} =k(0,c,\sigma
_{t},\alpha ).\qed
\end{align*}
}

\subsubsection{{\ {\ {Proof of \eqref{eqq:2}}}}}

{\small By Lemma \ref{lem:time-change}, $V(0, x^{l},x^{r}, c\lambda ,\sigma
_{0},\alpha )$ equals 
\begin{align*}
&=\lambda ^{-1}\sup_{\tau ^{\prime }}\mathbb{E}\left[ \max \left\{ \lambda
x^{l}+\int_{0}^{\tau ^{\prime }}\frac{\alpha ^{-1}}{\sigma
_{0}^{-2}+s\lambda ^{-2}\alpha ^{-2}}\mathrm{d}M_{s}^{l},\lambda
x^{r}+\int_{0}^{\tau ^{\prime }}\frac{\alpha ^{-1}}{\sigma
_{0}^{-2}+s\lambda ^{-2}\alpha ^{-2}}\mathrm{d}M_{s}^{r}\right\} -c\tau %
\right] \\
& =\lambda \sup_{\tau ^{\prime }}\mathbb{E}\left[ \max \left\{ \lambda
^{-1}x^{l}+\int_{0}^{\tau ^{\prime }}\frac{\alpha ^{-1}}{\sigma
_{0}^{-2}\lambda ^{2}+s\alpha ^{-2}}\mathrm{d}M_{s}^{l},\lambda
^{-1}x^{r}+\int_{0}^{\tau ^{\prime }}\frac{\alpha ^{-1}}{\sigma
_{0}^{-2}\lambda ^{2}+s\alpha ^{-2}}\mathrm{d}M_{s}^{r}\right\} -c\lambda
^{-3}\tau \right] \\
& =\lambda V(0,x_{1}\lambda ^{-1},x_{2}\lambda ^{-1},c\lambda ^{-3},\sigma
_{0}/\lambda ,\alpha ).
\end{align*}
Thus, 
\begin{eqnarray*}
k(0,c,\sigma _{0},\alpha ) &=&\inf \{x>0\colon 0=V(0,0,-x,c,\sigma
_{0},\alpha )\} \\
&=&\inf \{x>0\colon 0=V(0,0,-x\,\lambda ^{-1},c\,\lambda ^{-3},\sigma
_{0}/\lambda ,\alpha )\} \\
&=&\lambda \,\inf \{y>0\colon 0=V(0,0,-y,c\,\lambda ^{-3},\sigma
_{0}/\lambda ,\alpha )\} = \lambda \,k(0,c\,\lambda ^{-3},\sigma
_{0}/\lambda ,\alpha ).
\end{eqnarray*}%
Setting $\tilde{\sigma}_{0}=\sigma _{0}/\lambda $ gives the result.\qed }

\subsubsection{{\ {\ {Proof of \eqref{eqq:4}}}}}

{\small First, observe that $V(t,x_{1},x_{2},c,\sigma _{0},\lambda \alpha )$
equals 
\begin{align*}
& =\sup_{\tau\geq t}\mathbb{E}\left[ \max \left\{ x_{1}+\int_{t}^{\tau }\frac{%
\lambda ^{-1}\alpha ^{-1}}{\sigma _{0}^{-2}+\lambda^{-2} \alpha ^{-2}s}%
\mathrm{d}W_{s}^{1},x_{2}+\int_{t}^{\tau }\frac{\lambda ^{-1}\alpha ^{-1}}{%
\sigma _{0}^{-2}+\lambda^{-2} \alpha ^{-2}s}\mathrm{d}W_{s}^{2}\right\}
-c(\tau-t)\right] \\
& =\lambda \sup_{\tau\geq t}\mathbb{E}\left[ \max \left\{ \lambda^{-1}
x_{1}+\int_{t}^{\tau }\frac{\alpha ^{-1}}{\lambda^2\sigma_{0}^{-2}+%
\alpha^{-2}s}\mathrm{d}W_{s}^{1},\lambda^{-1} x_{2}+\int_{t}^{\tau }\frac{%
\alpha ^{-1}}{\lambda^2\sigma_{0}^{-2}+\alpha^{-2}s}\mathrm{d}%
W_{s}^{2}\right\} -(c\lambda^{-1} )(\tau-t)\right] \\
& =\lambda V(t,\lambda^{-1} x_{1},\lambda^{-1} x_{2},\lambda^{-1}
c,\lambda^{-1} \sigma _{0},\alpha )\,.
\end{align*}%
Thus, 
\begin{eqnarray*}
k(t,c,\sigma _{0},\lambda \alpha ) &=&\inf \{x>0\colon 0=V(t,0,-x,c,\sigma
_{0},\lambda \alpha )\} \\
&=&\inf \{x>0\colon 0=V(t,0,-\lambda^{-1} x,\lambda^{-1} c,\lambda^{-1}
\sigma _{0},\alpha )\} \\
&=&\lambda \,\inf \{y>0\colon 0=V(t,0,-y,\lambda^{-1} c,\lambda^{-1} \sigma
_{0},\alpha )\} = \lambda\,k(t,\lambda^{-1} c,\lambda^{-1} \sigma
_{0},\alpha ).\qed
\end{eqnarray*}
}

\subsubsection{{{{Proof of \eqref{eqq:3}}}}}

{\small By Lemma \ref{lem:time-change}, $V(0,x^l,x^r,c\lambda,\sigma_{0},
\alpha)$ equals {\footnotesize 
\begin{align*}
& \lambda^{-1}\sup_{\tau^{\prime }}\mathbb{E}\left[\max\left\{ \lambda
x^l+\int_{0}^{\tau^{\prime }}\frac{\alpha^{-1}}{\sigma_{0}^{-2}+s%
\lambda^{-2}\alpha^{-2}}\mathrm{d}M_{s}^{l},\lambda
x^r+\int_{0}^{\tau^{\prime }}\frac{\alpha^{-1}}{\sigma_{0}^{-2}+s%
\lambda^{-2}\alpha^{-2}}\mathrm{d}M_{s}^{r}\right\} -c\tau\right]
\end{align*}
As receiving more information is always better we have that for all $%
\lambda>1$ } {\footnotesize 
\begin{eqnarray*}
V(0,x^l,x^r,c\lambda,\sigma_{0}, \alpha) & \geq & \lambda^{-1}\sup_{\tau}%
\mathbb{E}\left[\max\left\{ \lambda x^l+\int_{0}^{\tau}\frac{\alpha^{-1}}{%
\sigma_{0}^{-2}+s\alpha^{-2}}\mathrm{d}M_{s}^{l},\lambda x^r+\int_{0}^{\tau}%
\frac{\alpha^{-1}}{\sigma_{0}^{-2}+s\alpha^{-2}}\mathrm{d}M_{s}^{r}\right\}
-c\tau\right] \\
& = & \lambda^{-1}V(0,\lambda x^l,\lambda x^r,c,\sigma_{0}, \alpha)
\end{eqnarray*}%
} This implies that $k(t,\lambda c, \sigma_0,
\alpha)\geq\lambda^{-1}k(t,c,\sigma_0, \alpha)$ for all $\lambda>1$ 
\begin{eqnarray*}
k(t,\lambda c, \sigma_0, \alpha) & = & \inf\{x>0\colon0=V(t,0,-x,\lambda c,\sigma_{0}, \alpha)\} \\
& \geq & \inf\{x>0\colon0=V(t,0,-x\lambda,c,\sigma_{0}, \alpha)\} \\
& = & \lambda^{-1}\inf\{y>0\colon0=V(0,0,-y,c,\sigma_{0}, \alpha)\} =
\lambda^{-1} k(t,c, \sigma_0, \alpha)\,. \qed
\end{eqnarray*}
}

\subsubsection{ Lipschitz continuity of $k$}

{\small Let $\lambda _{\epsilon }=\left( 1+\epsilon \alpha^{-2}\sigma
_{0}^{2}\right) ^{-1/2}<1$ and note that by definition $\lambda _{\epsilon
}\,\sigma _{0}=\sigma _{\epsilon }$. We can thus use equations \eqref{eqq:1}%
, \eqref{eqq:2} and \eqref{eqq:3} to get 
\begin{equation*}
k(\epsilon ,c,\sigma _{0},\alpha )=k(0,c,\sigma _{\epsilon },\alpha
)=\lambda _{\epsilon }k(0,c\lambda _{\epsilon }^{-3},\sigma _{0})\geq
\lambda _{\epsilon }^{4}k(0,c,\sigma _{0},\alpha )\,.
\end{equation*}%
As a consequence we can bound the difference between the value of the
barrier at time zero and at time $\epsilon $ from below 
\begin{equation*}
k(\epsilon ,c,\sigma _{0},\alpha )-k(0,c,\sigma _{0},\alpha )\geq \left[
\left( 1+\epsilon \alpha^{-2}\sigma _{0}^{2}\right) ^{-2}-1\right]
\,k(0,c,\sigma _{0},\alpha ).
\end{equation*}%
Dividing by $\epsilon $ and taking the limit $\epsilon \rightarrow 0$ yields
that $k_{t}$ the partial derivative of the boundary with respect to time
satisfies 
\begin{equation*}
k_{t}(0,c,\sigma _{0},\alpha )\geq -2\alpha^{-2}\sigma _{0}^{2} k(0,c,\sigma
_{0},\alpha )\,.
\end{equation*}%
Since by equation \eqref{eqq:1} $k(t+\epsilon ,c,\sigma _{0},\alpha
)=k(\epsilon ,c,\sigma _{t},\alpha )$ we have that {\footnotesize 
\begin{align*}
k_{t}(t,c,\sigma _{0},\alpha )&= \lim_{\epsilon \to 0}\frac{k(t+\epsilon
,c,\sigma _{0},\alpha ) - k(t,c,\sigma _{0},\alpha )}{\epsilon} =
\lim_{\epsilon \to 0}\frac{k(\epsilon ,c,\sigma _{t},\alpha ) - k(0,c,\sigma
_{t},\alpha )}{\epsilon} \\
&=k_{t}(0,c,\sigma _{t},\alpha )\geq -2\alpha^{-2}\sigma
_{t}^{2}k(0,c,\sigma _{t},\alpha )=-2\alpha^{-2}\sigma _{t}^{2}k(t,c,\sigma
_{0},\alpha )\geq -2\alpha^{-2}\sigma _{0}^{2}k(0,c,\sigma _{0},\alpha ),
\end{align*}%
}where the last equality follows from equation \eqref{eqq:1} and the last
inequality follows since $k$ and $\sigma _{t}$ are decreasing in $t$. Thus, $%
k_{t}$ is bounded from below; since its upper bound is zero, $k$ is
Lipschitz continuous in $t$. \qed }

\subsection{{{\ Proof of Theorem \protect\ref{thm:monot2}}}}

{\small \ If the agent stops at time $t$ at the barrier $k(t,c,\sigma
_{0},\alpha ),$ his posterior belief is that the true states are normally
distributed with $\theta ^{i}\sim \mathcal{N}(X_{t}^{i},\sigma _{t})$.
Thus,the probability that the agent assigns to picking $l$ when $r$ is
optimal, conditional on stopping at $t$, is  
{\footnotesize 
\begin{eqnarray*}
\mathbb{P}\left[ \theta
^{l}<\theta ^{r}\,|\,X_{t}^{l}-X_{t}^{r}=k(t,c,\sigma _{0},\alpha )\right] &=&\mathbb{P}\left[ (\theta ^{l}-\theta ^{r})-(X_{t}^{l}-X_{t}^{r})\leq
-k(t,c,\sigma _{0},\alpha )\,\bigg|\,X_{t}^{l}-X_{t}^{r}=k(t,c,\sigma
_{0},\alpha )\right] \\
&=&\mathbb{P}\left[ \frac{(\theta ^{l}-\theta ^{r})-(X_{t}^{l}-X_{t}^{r})}{%
\sqrt{2}\sigma _{t}}\leq -\frac{k(t,c,\sigma _{0},\alpha )}{\sqrt{2}\sigma
_{t}}\,\bigg|\,X_{t}^{l}-X_{t}^{r}=k(t,c,\sigma _{0},\alpha )\right] \\
&=&\Phi \left( -\frac{1}{\sqrt{2}}k(t,c,\sigma _{0},\alpha )\sigma
_{t}^{-1}\right) \,.
\end{eqnarray*}%
} From the symmetry of the problem, there is the same probability of
mistakenly picking $r$ instead of $l$. To show that the probability of being
wrong increases over time, it remains to show that $k(t,c,\sigma _{0},\alpha
)\sigma _{t}^{-1}$ is decreasing in $t$. We have that 
\begin{equation*}
\frac{\partial }{\partial \sigma _{t}}\left[ k(0,\sigma _{t},c,\alpha
)\sigma _{t}^{-1}\right] =k_{\sigma }(0,c,\sigma _{t},\alpha )\sigma
_{t}^{-1}-k(0,c,\sigma _{t},\alpha )\sigma _{t}^{-2}.
\end{equation*}%
We will now show that this is equal to $-3\,c\,k_{c}(0,\sigma _{t},c,\alpha
)\sigma _{t}^{2}$, which is nonnegative. To see that, we show that $%
k_{\sigma }(0,c,\sigma _{0},\alpha )\sigma _{0}=-3\,c\,k_{c}(0,c,\sigma
_{0},\alpha )+k(0,c,\sigma _{0},\alpha )$. Set 
\begin{equation*}
\beta _{\epsilon }\sigma _{0}=\sigma _{0}+\epsilon \Rightarrow \beta
_{\epsilon }=1+\frac{\epsilon }{\sigma _{0}}\,.
\end{equation*}%
Inserting in equation. \eqref{eqq:2} gives 
\begin{eqnarray*}
k(0,c,\sigma _{0}\,\beta _{\epsilon },\alpha ) &=&k(0,c,\sigma _{0}+\epsilon
,\alpha )=\beta _{\epsilon }\,k(0,c\,\beta _{\epsilon }^{-3},\sigma
_{0},\alpha ) \\
\Leftrightarrow k(0,c,\sigma _{0}+\epsilon ,\alpha )-k(0,c,\sigma
_{0},\alpha ) &=&k(0,c\,\beta _{\epsilon }^{-3},\sigma _{0},\alpha
)-k(0,c,\sigma _{0},\alpha )+\frac{\epsilon }{\sigma _{0}}k(0,c\,\beta
_{\epsilon }^{-3},\sigma _{0},\alpha ).
\end{eqnarray*}%
Dividing by $\epsilon $ and taking the limit $\epsilon \rightarrow 0$ yields 
\begin{eqnarray*}
k_{\sigma }(0,c,\sigma _{0},\alpha ) &=&k_{c}(0,c,\sigma _{0},\alpha )c\left[
\lim_{\epsilon \rightarrow 0}\frac{\beta _{\epsilon }^{-3}-1}{\epsilon }%
\right] +\frac{1}{\sigma _{0}}k(0,c,\sigma _{0},\alpha )  \notag \\
&=&k_{c}(0,c,\sigma _{0},\alpha )c\left[ -3\right] \frac{\partial \beta
_{\epsilon }}{\partial \epsilon }+\frac{1}{\sigma _{0}}k(0,c,\sigma
_{0},\alpha )  \notag \\
&=&-k_{c}(0,c,\sigma _{0},\alpha )c\frac{3}{\sigma _{0}}+\frac{1}{\sigma _{0}%
}k(0,c,\sigma _{0},\alpha )  \notag \\
\Leftrightarrow k_{\sigma }(0,c,\sigma _{0},\alpha )\sigma _{0}
&=&-3\,c\,k_{c}(0,c,\sigma _{0},\alpha )+k(0,c,\sigma _{0},\alpha )\,.
\end{eqnarray*}%
}

\subsection{{\ {{\ {Proof of Fact \protect\ref{fact:Chernoff}}}}}}

{\small Let $\kappa:=\mathbb{E}\big[\max\{\theta^l, \theta^r\}\big]$ and fix
a stopping time $\tau$. To show that 
\begin{equation*}
\mathbb{E}\left[-\mathbf{1}_{\{X^l_\tau \geq X^r_\tau\}}
(\theta^r-\theta^l)^+ -\mathbf{1}_{\{X^r_\tau > X^l_\tau\}}
(\theta^l-\theta^r)^+ -c \tau \right] = \mathbb{E}\left[\max\{X^l_\tau,
X^r_\tau\}-c\tau \right] + \kappa,
\end{equation*}
the cost terms can be dropped. Let $D$ be the difference between the expected payoff from the optimal decision and the expected payoff from choosing the correct action,  $D:=\mathbb{E}\big[\max\{X^l_\tau, X^r_\tau\}\big]- \mathbb{E}\big[%
\max\{\theta^l, \theta^r\}\big]$. By decomposing the expectation into two events,
\begin{align*}
D&= \mathbb{E}\Big[ \mathbf{1}_{\{X^l_\tau \leq X^r_\tau \}} (X^l_\tau
-\max\{\theta^l, \theta^r\}) + \mathbf{1}_{\{X^r_\tau < X^l_\tau \}}
(X^r_\tau -\max\{\theta^l, \theta^r\}) \Big]. \\
\shortintertext{Plugging in the definition of $X^i_\tau$ and using the law of iterated expectations,}
D &= \mathbb{E}\bigg[ \mathbf{1}_{\{X^l_\tau \leq X^r_\tau \}} (\mathbb{E}%
[\theta^l|\mathcal{F}_\tau] -\max\{\theta^l, \theta^r\}) + \mathbf{1}%
_{\{X^r_\tau < X^l_\tau \}} (\mathbb{E}[\theta^r|\mathcal{F}_\tau]
-\max\{\theta^l, \theta^r\}) \bigg] \\
&= \mathbb{E}\bigg[ \mathbf{1}_{\{X^l_\tau \leq X^r_\tau \}} (\mathbb{E}%
[\theta^l|\mathcal{F}_\tau] -\mathbb{E}[\max\{\theta^l, \theta^r\}|%
\mathcal{F}_\tau]) + \mathbf{1}_{\{X^r_\tau < X^l_\tau \}} (\mathbb{E}%
[\theta^r|\mathcal{F}_\tau] -\mathbb{E}[\max\{\theta^l, \theta^r\}|%
\mathcal{F}_\tau] ) \bigg] \\
&= \mathbb{E}\bigg[ \mathbf{1}_{\{X^l_\tau \leq X^r_\tau \}} \mathbb{E}%
[-(\theta^r-\theta^l)^+|\mathcal{F}_\tau] + \mathbf{1}_{\{X^r_\tau
< X^l_\tau \}} \mathbb{E}[-(\theta^l - \theta^r)^+|\mathcal{F}%
_\tau] \bigg] \\
&= \mathbb{E}\bigg[- \mathbf{1}_{\{X^l_\tau \leq X^r_\tau \}}
(\theta^r-\theta^l)^+ - \mathbf{1}_{\{X^r_\tau < X^l_\tau
\}}(\theta^l - \theta^r)^+ \bigg].
\end{align*}
}

\subsection{{\ {\ {{Proof of Fact \protect\ref{fact:Bather}}}}}}
{\small
We rely on \citeauthor{Bather62}'s \citeyearpar{Bather62} analysis of
the Chernoff model, which by Fact \ref{fact:Chernoff} applies to our model.
Bather studies a model with zero prior precision. Since such an agent
never stops instantaneously, all that matters is his beliefs at $t>0$, which
are well defined even in this case, and given by $\hat X_{t}^{i}=t^{-1}Z_{t}^{i}$ and $\hat \sigma _{t}^{-2}=t\alpha ^{-2}$. In Section 6, p. 619 \cite{Bather62} shows that 
\begin{align*} 
k\big(t,c,\infty ,\frac{1}{\sqrt{2}}\big)\sqrt{t}&=\frac{1}{4\,c\,t^{3/2}}+O\left( \frac{%
1}{t^{3}}\right).
\intertext{which implies that}
k\big(t,c,\infty ,\frac{1}{\sqrt{2}}\big)&=\frac{1}{4\,c\,t^2}+O\left( \frac{%
1}{t^{7/2}}\right).
\end{align*}
Fix $\alpha >0$. By equation \eqref{eqq:4} we have $k(t,c,\infty ,\alpha
)=\alpha \sqrt{2}k(t,\frac{1}{\alpha \sqrt{2}}c,\infty ,\frac{1}{\sqrt{2}})$%
. Thus, 
\begin{equation*}
k\big(t,c,\infty ,\alpha \big)=\frac{1}{2\,c\alpha ^{-2}\,t^{2}}+O\left( 
\frac{1}{t^{7/2}}\right) .
\end{equation*}%
This implies that there exists $T,\beta >0$ such that for all $t>T$ we have 
\begin{equation*}  
\left\vert k\big(t,c,\infty ,\alpha \big)-\frac{1}{2\,c\alpha ^{-2}\,t^{2}}%
\right\vert \leq \frac{\beta }{t^{7/2}}.
\end{equation*}

Fix $\sigma _{0}>0$ and let $s:=t-\alpha ^{2}\sigma _{0}^{-2}$. This way, the agent who starts with zero prior precision and waits $t$ seconds has the same posterior precision as the agent who starts with $\sigma_0^2$ and waits $s$ seconds.\footnote{To see this, observe that $%
\sigma_{s}^{2}=\frac{1}{\sigma_0^{-2}+\alpha^{-2}s^2}=
\frac{1}{\alpha ^{-2}t}=\hat \sigma_t^2$.}
Thus, by \eqref{eqq:1} we have $k\big(t,c,\infty ,\alpha \big)%
=k\big(s,c,\sigma _{0},\alpha \big)$, so 
\begin{equation*}
\left\vert k\big(s,c,\sigma _{0},\alpha \big)-\frac{1}{2\,c\alpha
^{2}\,(\alpha ^{-2}s+\sigma _{0}^{-2})^{2}}\right\vert \leq \frac{\beta }{(s+\alpha ^{2}\sigma _{0}^{-2})^{7/2}}.
\end{equation*}%
Finally, since $b(s,c,\sigma _{0},\alpha )=k(s,c,\sigma _{0},\alpha )\sigma
_{s}^{-2}$, we have 
\begin{equation*}
\left\vert b\big(s,c,\sigma _{0},\alpha \big)-\frac{1}{2\,c\alpha
^{2}\,(\alpha ^{-2}s+\sigma _{0}^{-2})}\right\vert \leq \frac{\beta }{(s+\alpha ^{2}\sigma _{0}^{-2})^{5/2}}.
\end{equation*}}

{\small To see that \eqref{eq:k} and \eqref{eq:b} hold, notice that by equations \eqref{eqq:1}, \eqref{eqq:4}, \eqref{eqq:2} and %
\eqref{eqq:3} applied in that order, it follows that }%
\begin{align*}
\bar{k}(t,c,\sigma _{0},\alpha )& =\bar{k}(0,c,\sigma _{t},\alpha )=\alpha 
\bar{k}(0,\alpha ^{-1}c,\alpha ^{-1}\sigma _{t},1)=\sigma _{t}\bar{k}%
(0,\alpha ^{2}c\sigma _{t}^{-3},1,1) \\
& =\alpha ^{-2}c^{-1}\sigma _{t}^{4}\bar{k}(0,1,1,1)=\frac{\kappa }{c\alpha
^{2}(\sigma _{0}^{-2}+\alpha ^{-2}t)^{2}},
\end{align*}%
where $\kappa =\bar{k}(0,1,1,1)$. Since $\bar{b}(t,c,\sigma _{0},\alpha )=%
\bar{k}(t,c,\sigma _{0},\alpha )\sigma _{t}^{-2}$, it follows that $\bar{b}%
(t,c,\sigma _{0},\alpha )=\frac{\kappa }{c\alpha ^{2}(\sigma
_{0}^{-2}+\alpha ^{-2}t)}$. The fact that $\kappa =\frac{1}{2}$ follows from
the proof of Fact \ref{fact:Bather}, as any other constant would result in a
contradiction as $t\rightarrow \infty $.

\subsection{{Proof of Theorem \protect\ref{thm-implementability}}}

{\small Let $G=\{t_n\}_{n=1}^N$ be a finite set of times at which the agent
is allowed to stop and denote by $\mathcal{T}$ all stopping times $\tau$
such that $\tau \in G$ almost surely. As we restrict the agent to stopping
times in $\mathcal{T}$, the stopping problem becomes a discrete time optimal
stopping problem. By Doob's optional sampling theorem we have that 
\begin{align*}
\sup_{\tau }\mathbb{E}\left[ \max \{X_{\tau }^{l},X_{\tau }^{r}\}-d(\tau)%
\right] & =\sup_{\tau }\mathbb{E}\left[ \frac{1}{2}\max \{X_{\tau
}^{l}-X_{\tau }^{r},X_{\tau }^{r}-X_{\tau }^{l}\}+\frac{1}{2}(X_{\tau
}^{l}+X_{\tau }^{r})-d(\tau)\right] \\
& =\sup_{\tau }\mathbb{E}\left[ \frac{1}{2}|X_{\tau }^{l}-X_{\tau
}^{r}|-d(\tau)\right] +\frac{1}{2}(X_{0}^{l}+X_{0}^{r}),
\end{align*}%
so any optimal stopping time also solves $\sup_{\tau }\mathbb{E}\left[
|X_{\tau }^{l}-X_{\tau }^{r}|-2\,d(\tau)\right] $. 
Define $\Delta _{n}=|X_{t_{n}}^{l}-X_{t_{n}}^{r}|$ for all $n=1, \ldots, N$.
Observe that $(\Delta _{n})_{n=1, \ldots, N}$ is a one-dimensional discrete
time Markov process. To prove that for every barrier there exists a cost
function which generates $b$ by Theorem 1 in \cite{kruse2014optimal} it
suffices to prove that: }

\begin{enumerate}
\item {\small there exists a constant $C$ such that $\mathbb{E}[\Delta
_{n+1}|\mathcal{F}_{t_{n}}]\leq C(1+\Delta _{n})$ . }

\item {\small $\Delta_{n+1}$ is increasing in $\Delta_n$ in the sense of
first order stochastic dominance }

\item {\small $z(n,y)=\mathbb{E}[\Delta _{n+1}-\Delta _{n}\mid \Delta
_{n}=y] $ is strictly decreasing in $y.$ }
\end{enumerate}

{\small Condition 1 keeps the value of continuing from exploding, which
would be inconsistent with a finite boundary. Conditions 2 and 3 combined
ensure that the optimal policy is a cut-off rule. }

\subsubsection{Certain-Difference DDM}

{\small Set $Z_{t}=Z_{t}^{l}-Z_{t}^{r}=(\theta ^{\prime \prime}
-\theta^{\prime })t+\sqrt{2}\alpha B_{t}$. Then 
\begin{align*}
l_{t}& =\log \left( \frac{\mathbb{P}\left[ \theta =\theta _{l}\,|\,\mathcal{F%
}_{t}\right] }{\mathbb{P}\left[ \theta =\theta _{r}\,|\,\mathcal{F}_{t}%
\right] }\right) =\log \left( \frac{\mu }{1-\mu }\right) +\log \left( \frac{%
\exp (-(4\alpha ^{2}t)^{-1})(Z_{t}-(\theta ^{\prime \prime }-\theta ^{\prime
})t)^{2}}{\exp (-(4\alpha ^{2}t)^{-1})(Z_{t}-(\theta ^{\prime }-\theta
^{\prime \prime })t)^{2}}\right) \\
& =\log \left( \frac{\mu }{1-\mu }\right) +\frac{Z_{t}(\theta ^{\prime
\prime }-\theta ^{\prime })}{\alpha ^{2}}.
\end{align*}
}

{\small Denote by $p_{n}=\mathbb{P}[\theta =\theta _{l}\mid \mathcal{F}%
_{t_{n}}]$ the posterior probability that $l$ is the better choice. The
expected absolute difference of the two choices satisfies 
\begin{align*}
\Delta _{n}& =|X_{t_{n}}^{l}-X_{t_{n}}^{r}|=|p_{n}(\theta ^{\prime \prime
}-\theta ^{\prime })+(1-p_{n})(\theta ^{\prime }-\theta ^{\prime \prime })|
\\
& =|(2p_{n}-1)(\theta ^{\prime \prime }-\theta ^{\prime })|=2(\theta
^{\prime \prime }-\theta ^{\prime })\,\,\left\vert p_{n}-\frac{1}{2}%
\right\vert \,.
\end{align*}
Let $\psi
_{n}:=[Z_{t_{n}}^{l}-Z_{t_{n}}^{r}]-[Z_{t_{n-1}}^{l}-Z_{t_{n-1}}^{r}]$
denote the change in the signal from $t_{n-1}$ to $t_{n}$. We have that the
log likelihood is given by $l_{n+1}=l_{n}+\alpha ^{-2}(\theta ^{\prime
\prime }-\theta ^{\prime })\psi _{n+1}$. We thus have 
\begin{equation}
\Delta _{n}=2(\theta ^{\prime \prime }-\theta ^{\prime })\,\,\left\vert 
\frac{e^{l_{n}}}{1+e^{l_{n}}}-\frac{1}{2}\right\vert =2(\theta ^{\prime
\prime }-\theta ^{\prime })\left( \frac{e^{|l_{n}|}}{1+e^{|l_{n}|}}-\frac{1}{%
2}\right) \,.  \label{eq:folded-normal-one}
\end{equation}%
\noindent (1): It is easily seen that $\mathbb{E}\left[ \Delta _{n+1}|%
\mathcal{F}_{t_{n}}\right] \leq (\theta ^{\prime \prime }-\theta ^{\prime })$%
, so for $C$ big enough, $\mathbb{E}[\Delta _{n+1}|\mathcal{F}_{t_{n}}]\leq
C(1+\Delta _{n})$. \newline
(2): To simplify notation we introduce $m_{n}=|l_{n}|$. The process $%
(m_{n})_{n=1,\ldots ,N}$ is Markov. More precisely, $m_{n+1}=|m_{n}+\alpha
^{-2}(\theta ^{\prime \prime }-\theta ^{\prime })\psi _{n+1}|$ \ is folded
normal with mean of the underlying normal distribution equal to 
\begin{align}
m_{n}+\alpha ^{-2}(\theta ^{\prime \prime }-\theta ^{\prime })\mathbb{E}%
[\psi _{n+1}|l_{n}]& =|l_{n}|+\alpha ^{-2}(\theta ^{\prime \prime }-\theta
^{\prime })\Delta _{n}(t_{n+1}-t_{n})  \notag \\
& =m_{n}+\alpha ^{-2}\Big(\frac{2e^{m_{n}}}{1+e^{m_{n}}}-1\Big)(\theta
^{\prime \prime }-\theta ^{\prime })^{2}(t_{n+1}-t_{n})
\label{eq:folded-normal-two}
\end{align}%
and variance 
\begin{align*}
Var\left[ m_{n}+\alpha ^{-2}(\theta ^{\prime \prime }-\theta ^{\prime })\psi
_{n+1}\right] & =\alpha ^{-4}(\theta ^{\prime \prime }-\theta ^{\prime
})^{2}Var\left[ \psi _{n+1}\right] \\
& =2\alpha ^{-4}(\theta ^{\prime \prime }-\theta ^{\prime
})^{2}(t_{n+1}-t_{n})\,.
\end{align*}
As argued in part (2) of the uncertain difference case, a folded normal
random variable increases in the sense of first order stochastic dominance
in the mean of the underlying normal distribution. As %
\eqref{eq:folded-normal-two} increases in $m_{n}$ it follows that $m_{n+1}$
increases in $m_{n}$ in the sense of first order stochastic dominance. By %
\eqref{eq:folded-normal-one} $m_{n}=|l_{n}|$ is increasing in $\Delta _{n}$
and $\Delta _{n+1}$ is increasing in $m_{n+1}=|l_{n+1}|$ this completes the
argument.\newline
(3): It remains to show that $z(n,\Delta _{n})$ is decreasing in $\Delta
_{n} $. As $(p_{n})_{n=1,\ldots ,M}$ is a martingale, and moreover
conditioning on $p$ is equivalent to conditioning on $1-p,$ we have that 
\begin{align*}
z(n,\Delta _{n})& =\mathbb{E}\left[ \Delta _{n+1}\,\Big|\,p_{i}=\frac{\Delta
_{i}}{2(\theta ^{\prime \prime }-\theta ^{\prime })}+\frac{1}{2}\right]
-\Delta _{n} \\
& =2(\theta ^{\prime \prime }-\theta ^{\prime })\mathbb{E}\left[ |p_{n+1}-%
\frac{1}{2}|\,\Big|\,p_{n}=\frac{\Delta _{n}}{2(\theta ^{\prime \prime
}-\theta ^{\prime })}+\frac{1}{2}\right] -\Delta _{n} \\
& =2(\theta ^{\prime \prime }-\theta ^{\prime })\mathbb{E}\left[ p_{n+1}-%
\frac{1}{2}\,\Big|\,p_{i}=\frac{\Delta _{i}}{2(\theta ^{\prime \prime
}-\theta ^{\prime })}+\frac{1}{2}\right] \\
& \hspace{3cm}+2(\theta "-\theta ^{\prime })\mathbb{E}\left[ 2\max \Big\{{%
\frac{1}{2}-p_{n+1}},0\Big\}\,\Big|\,p_{i}=\frac{\Delta _{n}}{(\theta
^{\prime \prime }-\theta ^{\prime })}+\frac{1}{2}\right] -\Delta _{n}\,.
\end{align*}%
As $p$ is a martingale we can replace $p_{n+1}$ by $p_{n}$ 
\begin{align*}
z(n,\Delta _{n})& =2(\theta ^{\prime \prime }-\theta ^{\prime })\mathbb{E}%
\left[ p_{n}-\frac{1}{2}\,\Big|\,p_{n}=\frac{\Delta _{n}}{2(\theta ^{\prime
\prime }-\theta ^{\prime })}+\frac{1}{2}\right] \\
& \hspace{3cm}+2(\theta ^{\prime \prime }-\theta ^{\prime })\mathbb{E}\left[
2\max \left\{ {\frac{1}{2}-p_{n+1}},0\right\} \,\Big|\,p_{i}=\frac{\Delta
_{n}}{2(\theta ^{\prime \prime }-\theta ^{\prime })}+\frac{1}{2}\right]
-\Delta _{n} \\
& =\Delta _{n}+2(\theta ^{\prime \prime }-\theta ^{\prime }))\mathbb{E}\left[
2\max \left\{ {\frac{1}{2}-p_{n+1}},0\right\} \,\Big|\,p_{n}=\frac{\Delta
_{n}}{2(\theta ^{\prime \prime }-\theta ^{\prime })}+\frac{1}{2}\right]
-\Delta _{n} \\
& =4(\theta ^{\prime \prime }-\theta ^{\prime })\mathbb{E}\left[ \max
\left\{ {\frac{1}{2}-p_{n+1}},0\right\} \,\Big|\,p_{n}=\frac{\Delta _{n}}{%
2(\theta ^{\prime \prime }-\theta ^{\prime })}+\frac{1}{2}\right] \,.
\end{align*}%
The above term is strictly decreasing in $\Delta _{n}$ as $p_{n+1}$
increases in the sense of first order stochastic dominance in $p_{n}$ and $%
p_{n}$ in the conditional expectation is increasing in $\Delta _{n}$. }

\subsubsection{Uncertain-Difference DDM}

{\small Let us further define $\beta _{i}^{2}=2\sigma _{t_{i}}^{2}-2\sigma
_{t_{i+1}}^{2}$. As $X_{t_{i}+1}^{l}-X_{t_{i}+1}^{r}$ is Normal distributed
with variance $\beta _{i}^{2}$ and mean $\Delta _{i}$ we have that $\Delta
_{i+1}$ is folded normal distributed with mean 
\begin{equation*}
\mathbb{E}_{i}\left[ \Delta _{i+1}\right] =\beta _{i}\sqrt{\frac{2}{\pi }}%
e^{-\frac{\Delta _{i}^{2}}{2\beta _{i}^{2}}}+\Delta _{i}(1-2\Phi (\frac{%
-\Delta _{i}}{\beta _{i}}))\,,
\end{equation*}%
where $\Phi $ denotes the normal cdf. Thus, the expected change in delta is
given by 
\begin{equation*}
z(i,y)=\beta _{i}\sqrt{\frac{2}{\pi }}e^{-\frac{y^{2}}{2\beta _{i}^{2}}%
}-2\,y\,\Phi (\frac{-\Delta _{i}}{\beta _{i}}).
\end{equation*}
}

{\small \noindent (1): It is easily seen that $\mathbb{E}_i \left[
\Delta_{i+1}\right] \leq \beta_i \sqrt{\frac{2}{\pi}} + \Delta_i$.\newline
(2): As $\Delta_i$ is folded normal distributed we have that 
\begin{equation*}
\mathbb{P}_i (\Delta_{i+1} \leq y) = \frac{1}{2} \left[ \text{erf}\left( \frac{%
y+\Delta_i}{\beta_i}\right) + \text{erf}\left( \frac{y-\Delta_i}{\beta_i}\right)%
\right] \,.
\end{equation*}
Taking derivatives gives that 
\begin{align*}
\frac{\partial}{\partial \Delta_i} \mathbb{P}_i (\Delta_{i+1} \leq y)&=\frac{%
1}{2} \left[ e^{ -\left( \frac{y+\Delta_i}{\beta_i}\right)^2} - e^{-\left( 
\frac{y-\Delta_i}{\beta_i}\right)^2}\right] = \frac{1}{2} e^{-\left( \frac{y-\Delta_i}{\beta_i}\right)^2} \left[ e^{ -%
\frac{4 \Delta_i y}{\beta_i}} - 1\right] < 0 \,.
\end{align*}
As $\Delta_i = |X_{t_{i}}^{l}-X_{t_{i}}^{r}|$ it follows that $y\geq 0$ and
hence, $\Delta_{i+1}$ is increasing in $\Delta_i$ in the sense of first
order stochastic dominance.\newline
(3): The derivative of the expected change of the process $\Delta$ equals 
\begin{align*}
\frac{\partial}{\partial y}z(i,y) &= \frac{\partial}{\partial y}
\left(\beta_i \sqrt{\frac{2}{\pi}} e^{-\frac{y^2}{2 \beta_i^2}} - 2 \, y
\,\Phi(\frac{-\Delta_i}{\beta_i})) \right) = - 2 \Phi(\frac{-y}{\beta_i}) <
0 \, .
\end{align*}
Hence, $z$ is strictly decreasing in $y$.\qed
}

{\small 
}

{\small 
\bibliographystyle{econometrica}
\bibliography{DDM}
}

\end{document}